\documentclass[preprint,secnumarabic,amssymb, amsmath, nobibnotes, aps, prl, floatfix]{revtex4-2}

\usepackage{graphicx}
\usepackage[dvipsnames]{xcolor}
\usepackage{cleveref}
\usepackage{etoolbox}
\usepackage{tikz}
\usetikzlibrary{arrows.meta}
\usetikzlibrary{shapes.geometric}
\usepackage{pgfplots}
\usepackage{marginnote} 
\usepackage[inline]{enumitem}

\begin{document}

\newcommand{\lb}[1]{{\color{red}#1}}
\newcommand{\LB}[1]{[{\color{magenta}#1}]}
\newcommand{\SC}[1]{[{\color{green}\textbf{#1}}]}
\newcommand{\MC}[1]{[{\color{blue}#1}]}
\newcommand{\wel}{$We_{\mathcal{L}}$}
\newcommand{\mur}{$\mu_d/\mu_c$}
\newcommand{\lra}[1]{\langle #1 \rangle }
\definecolor{lightblue}{RGB}{173,216,230} 

\title{
The interaction of droplet dynamics and turbulence cascade
}%

\author{Marco Crialesi-Esposito}%
\email[]{crialesi@to.infn.it}
\affiliation{FLOW, Department of Engineering Mechanics, KTH Royal Institute of Technology, Stockholm, Sweden}
\affiliation{INFN, Sezione di Torino, Via Pietro Giuria 1, 10125 Torino, Italy}

\author{Sergio Chibbaro}%
\email[ ]{sergio.chibbaro@upsaclay.fr}
\affiliation{ Universit\'e Paris-Saclay, CNRS, LISN, 91400 Orsay, France}
\affiliation{ SPEC, CEA, CNRS UMR 3680, Université Paris-Saclay, CEA Saclay, Gif-sur-Yvette, France}

\author{Luca Brandt}%
\email[]{luca@mech.kth.se}
\affiliation{FLOW, Department of Engineering Mechanics, KTH Royal Institute of Technology, Stockholm, Sweden}
\affiliation{Department of Energy and Process Engineering, Norwegian University of Science and Technology (NTNU), Trondheim, Norway}

\begin{abstract}
The dynamics of droplet fragmentation in turbulence is described in the Kolmogorov-Hinze framework. Yet, a quantitative theory is lacking at higher concentrations when strong interactions between the phases and coalescence become relevant, which is common in most flows. Here, we address this issue through a fully-coupled numerical study of the droplet dynamics in a turbulent flow at high Reynolds number. By means of time-space spectral statistics, not currently accessible to experiments, we demonstrate that the characteristic scale of the process, the Hinze scale, can be precisely identified as the scale at which the net energy exchange due to capillarity is zero. Droplets larger than this scale preferentially break up absorbing energy from the flow; smaller droplets, instead, undergo rapid oscillations and tend to coalesce releasing energy to the flow. Further, we link the droplet-size-distribution with the probability distribution of the turbulent dissipation. This shows that key in the fragmentation process is the local flux of energy which dominates the process at large scales, vindicating its locality.
\end{abstract}

\maketitle

\definecolor{green}{rgb}{0,0.5,0}
\definecolor{black}{rgb}{0,0,0}
\definecolor{a003}{rgb} {0.0, 0.5, 0.0}
\definecolor{a006}{rgb} {1.0, 0.0, 0.0}
\definecolor{a01}{rgb}  {0.0, 0.0, 1.0}
\definecolor{a02}{rgb}  {0.75, 0.0, 0.75}
\definecolor{a05}{rgb}  {1.0, 0.4980392156862745, 0.054901960784313725}

\definecolor{m001}{rgb} {0.5019607843137255, 0.5019607843137255, 0.0}
\definecolor{m01}{rgb}  {1.0, 0.0784313725490196, 0.5764705882352941}
\definecolor{m1}{rgb}   {0.0, 0.75, 0.75}
\definecolor{m10}{rgb}  {1.0, 0.8431372549019608, 0.0}
\definecolor{m100}{rgb} {0.0, 0.39215686274509803, 0.0}

\definecolor{fam_a}{rgb}     {1.0, 0.0, 0.0}
\definecolor{fam_we}{rgb}    {0.0, 0.5, 0.0}
\definecolor{fam_mu10}{rgb}  {0.75, 0.0, 0.75}
\definecolor{fam_mu3}{rgb}   {1.0, 0.4980392156862745, 0.054901960784313725}

\scalebox{0}{%
\begin{tikzpicture}
    \begin{axis}[hide axis]
     	\addplot [
    	color=black,
    	dashed,
    	line width=0.9pt,
    	forget plot
    	]
    	(0,0);\label{dashed}
    	
    	\addplot [
    	color=black,
    	line width=0.9pt,
    	forget plot
    	]
    	(0,0);\label{conti}
    	
        \addplot [
        color=a003,
        dashed,
        line width=0.9pt,
        forget plot
        ]
        (0,0);\label{alpha003}
        
        \addplot [
        color=a006,
        dashed,
        line width=0.9pt,
        forget plot
        ]
        (0,0);\label{alpha006}
        
        \addplot [
        color=a01,
        dashed,
        line width=0.9pt,
        forget plot
        ]
        (0,0);\label{alpha01}
        
        \addplot [
        color=a02,
        dashed,
        line width=0.9pt,
        forget plot
        ]
        (0,0);\label{alpha02}
        
        \addplot [
        color=a05,
        dashed,
        line width=0.9pt,
        forget plot
        ]
        (0,0);\label{alpha05}
        
        \addplot [
        color=m001,
        dashdotted,
        line width=0.9pt,
        forget plot
        ]
        (0,0);\label{m001}
        
        \addplot [
        color=m01,
        dashdotted,
        line width=0.9pt,
        forget plot
        ]
        (0,0);\label{m01}
        
        \addplot [
        color=m1,
        dashdotted,
        line width=0.9pt,
        forget plot
        ]
        (0,0);\label{m1}
        
        \addplot [
        color=m10,
        dashdotted,
        line width=0.9pt,
        forget plot
        ]
        (0,0);\label{m10}
        
        \addplot [
        color=m100,
        dashdotted,
        line width=0.9pt,
        forget plot
        ]
        (0,0);\label{m100}
        
        \addplot [
        only marks,
        color=fam_a,
        mark=*,
        opacity=0.5,
        forget plot
        ]
        (0,0);\label{fam_a}
        
        \addplot [
        only marks,
        color=fam_we,
        mark=*,
        opacity=0.5,
        forget plot
        ]
        (0,0);\label{fam_we}

        \addplot [
        only marks,
        color=fam_mu3,
        mark=*,
        opacity=0.5,
        forget plot
        ]
        (0,0);\label{fam_mu3}
        
        \addplot [
        only marks,
        color=fam_mu10,
        mark=*,
        opacity=0.5,
        forget plot
        ]
        (0,0);\label{fam_mu10}
        
    \end{axis}
\end{tikzpicture}%
}%
\section*{Introduction}
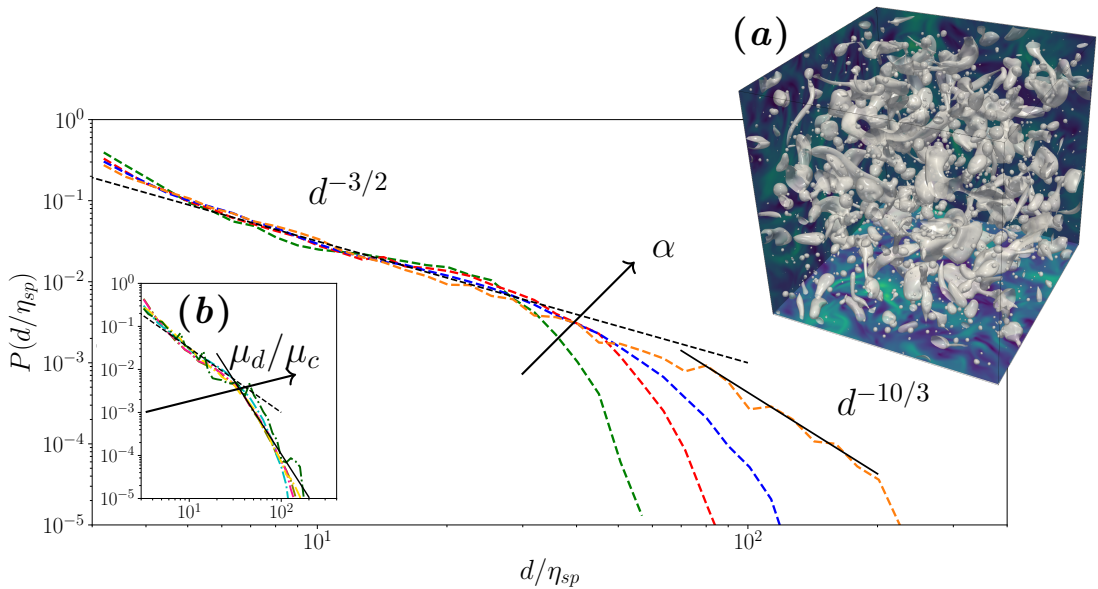
\begin{figure*}[tp]
	\centering
	\includegraphics[width=0.85\textwidth]{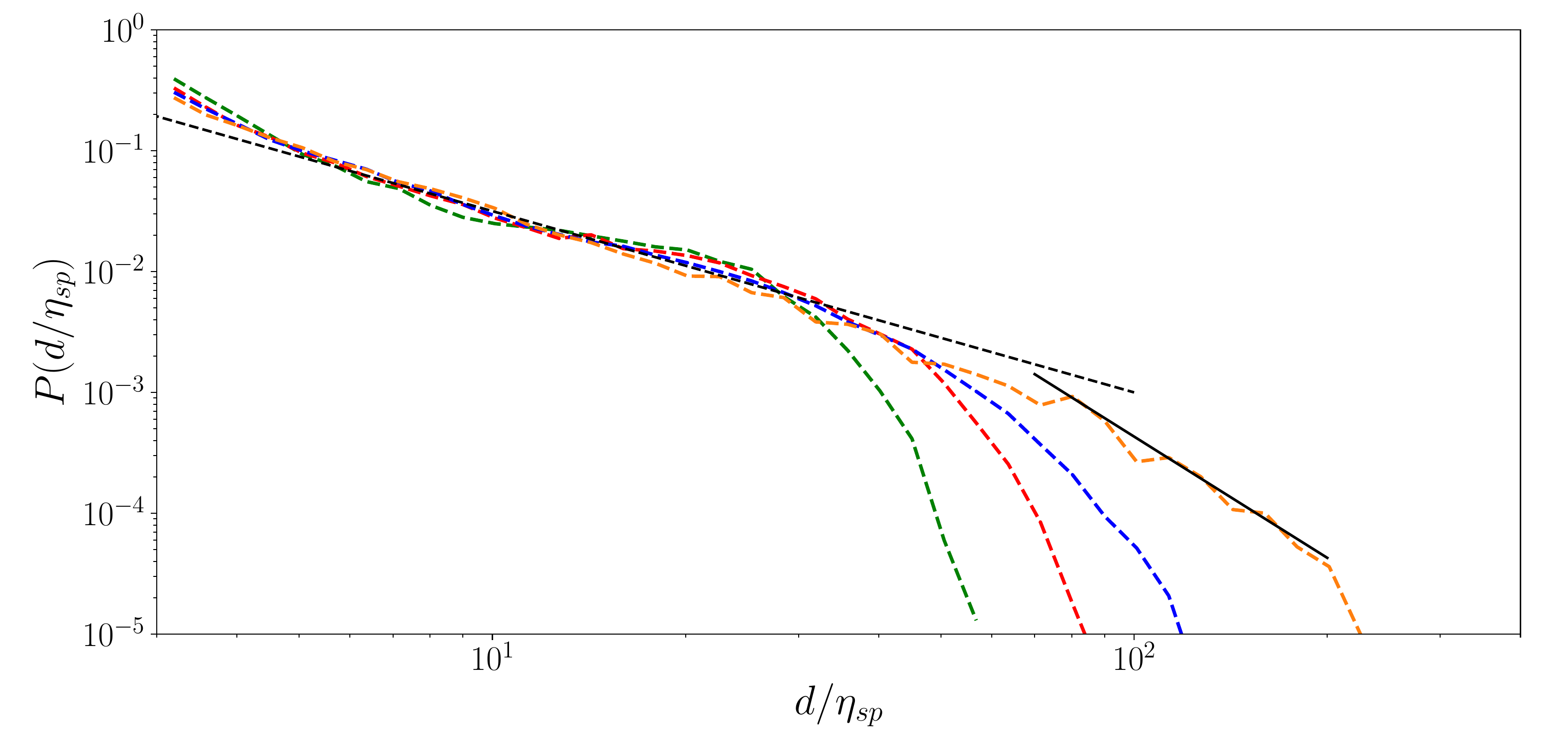}
	\begin{tikzpicture}[remember picture,overlay]
	\node[coordinate] (A) at (-7,3) {};
	\node[coordinate] (B) at (-5.5,4.5) {};
	\draw[thick,->] (A) -- (B);
	\end{tikzpicture}
	\put(-150,130){\large$\alpha$}
	\put(-280,150){{\large$d^{-3/2}$}}
	\put(-80,70){{\large$d^{-10/3}$}}
	\put(-360,30){\includegraphics[width=0.2\textwidth]{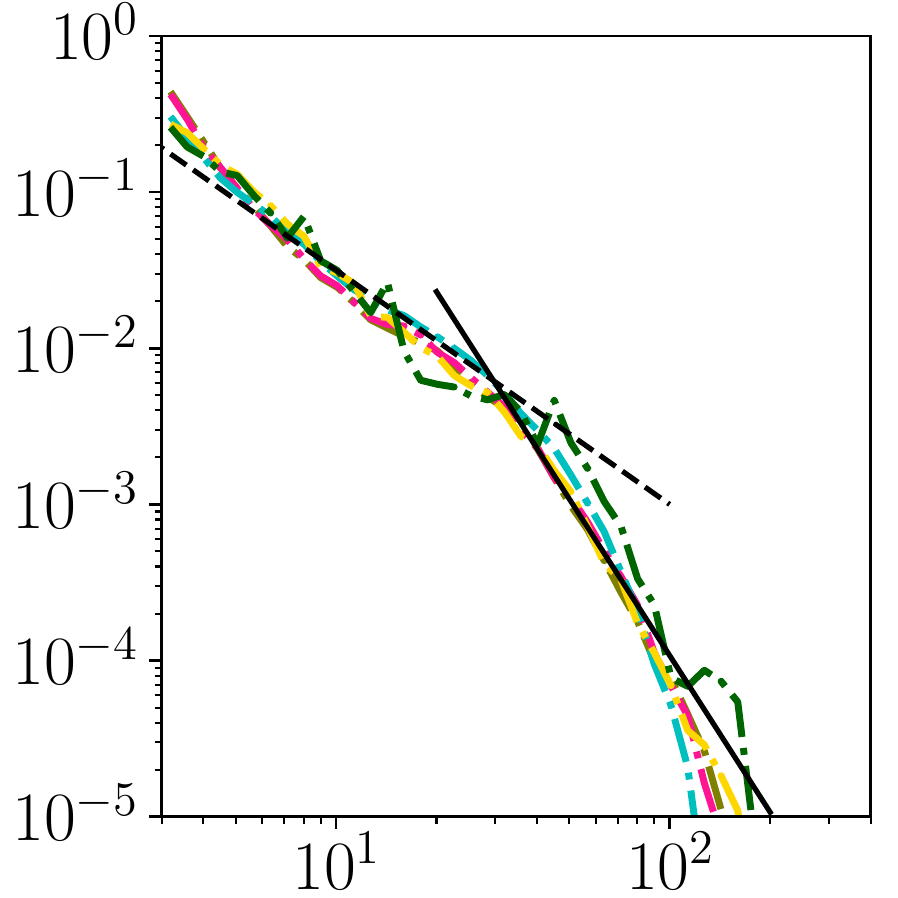}}
	\begin{tikzpicture}[remember picture,overlay]
	\node[coordinate] (A) at (-12,2.5) {};
	\node[coordinate] (B) at (-10,3) {};
	\draw[thick,->] (A) -- (B);
	\end{tikzpicture}
	\put(-310,90){\large$\mu_d/\mu_c$}
	\put(-330,105){\textbf{\large(\textit{b})}}
	\put(-120,70){\includegraphics[scale=0.1]{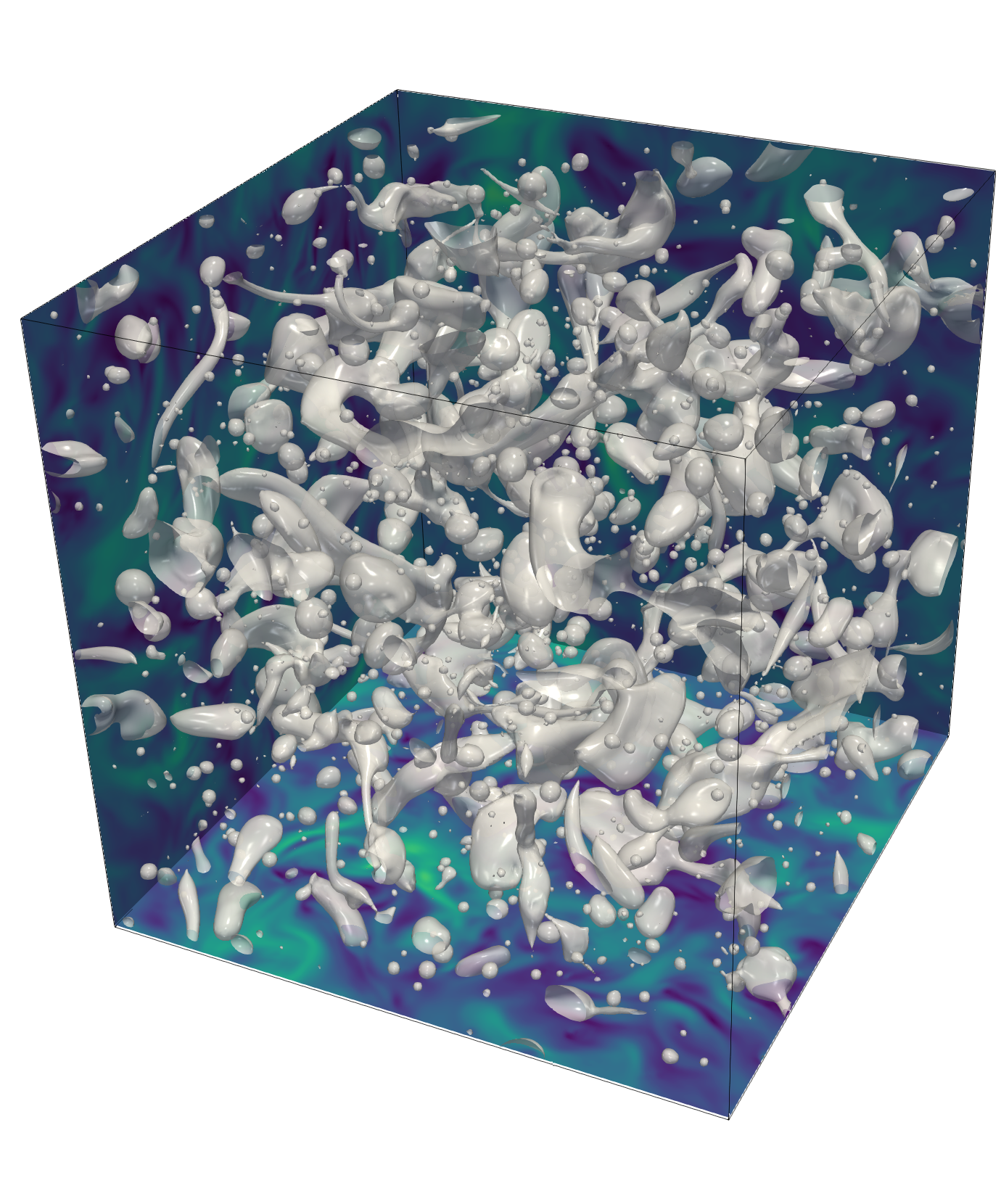}}
	\put(-120,210){\textbf{\large(\textit{a})}}
	\caption{
		Droplet-size-distribution for emulsions in HIT at different volume fractions $\alpha$, \emph{i.e.} 
		0.03 (\ref{alpha003})
		0.06 (\ref{alpha006})
		0.1  (\ref{alpha01})
		0.5  (\ref{alpha05}),
		at constant $We_\mathcal{L}=42.6$ and $\mu_d/\mu_c=1$. The droplet diameter is normalized with the Kolmogorov scale $\eta_{sp}$ for the single-phase reference case (also used as initial condition for all simulations) at $Re_\lambda=137$.
		Similarly, inset (\textit{b}) shows the DSD different viscosity ratio $\mu_d/\mu_c$, \emph{i.e.}
		0.01 (\ref{m001})
		0.1 (\ref{m01})
		1 (\ref{m1})
		10 (\ref{m10})
		100 (\ref{m100}), at constant $\alpha=0.1$ and $We_\mathcal{L}=42.6$. 
		The  dashed line  (\ref{dashed}) indicates the $-3/2$ power-law and the continuous line (\ref{conti}) the $-10/3$ power-law.
		Inset (\textit{a}) shows a render of the  simulation at $We_\mathcal{L}=42.6$, $\alpha=0.1$ and $\mu_d/\mu_c=1$. 
	}
	\label{fig:dsd_render}
\end{figure*}

Turbulent flows with dispersed interfaces are at the core of many transfer processes in gas-liquid (atomisation and sprays)\cite{Villermaux2009,Keshavarz2016,villermaux2020fragmentation} and liquid-liquid (emulsions) systems \cite{Perlekar2014,girotto2021,Bakhuis2021}. 
Notably, air bubbles are key for the gas transfer between ocean and atmosphere\cite{Garrett2000,Gao2021,Deike2022}, and in the aerosol production through bursting \cite{Berny2021,Jiang2022}.
Despite the numerous experimental and numerical studies 
\cite{Skartlien2013,Yu2019,Mukherjee2019,Perrard2020,Riviere2021,Yi2021}, 
the nature of the interactions between droplets of different sizes and turbulence is not yet clear. 
The presence of a deforming/breaking/coalescing interface couples the two phases in a non-trivial way, absorbing and distributing energy over the whole spectrum of scales. 
The Kolmogorov-Hinze (KH) theory \cite{Kolmogorov1949,Hinze1955} is the cornerstone of existing models and applications; this framework is based on the breakup of isolated droplets in turbulence and identifies the scale $d_H$ above which a droplet breaks up due to the local environment turbulence and below which surface tension forces are able to resist the action of the turbulent eddies.
This picture, based on breakup only, is incomplete and has been recently challenged \cite{qi2022fragmentation}.
Here, we
wish to go a step forward and provide a comprehensive explanation
by considering configurations in which turbulence is modulated by the dispersed phase and both coalescence and breakup occur.
We will use an original statistical approach analyzing the energy fluxes from fully resolved numerical simulations to unambiguously 
show that there exists a scale $d_H$ which separates two regimes--one statistically dominated by droplet coalescence and the other by breakup. 
We also show that intermittency at small scales significantly increases in multiphase turbulence and demonstrate 
how the extreme-event distribution can be inferred directly by the droplet size distribution.

The key observable is the distribution spectrum of the size of the intrusions, $\mathcal{N}(d)$, which gives the number of droplet at a given size $d$.
The dynamics for large diameters has been 
rationalised in the KH framework in terms of a local fragmentation process \cite{Garrett2000}. 
The idea is that a droplet breaks up whenever the pressure forces acting on its surface are larger than the cohesive force given by the surface tension.
In this picture, the {only} dimensionless parameter  is  the Weber number $We=\rho_c u_d^2 d/\sigma$, where $\rho_c$ is the carrier-phase density, $\sigma$ is the surface tension, and $u_d$ is 
the typical velocity at the scale of the droplet size, $d$. 
The disruptive inertial{-range velocity} fluctuations can initiate 
fragmentation above a critical  threshold $We_c$ \cite{fuster2021vortex}. 
Assuming the local Kolmogorov description of turbulence \cite{Kolmogorov1941,Garrett2000},  $ u_d^2 \sim \lra{\varepsilon}^{2/3} d^{2/3}$, where $\lra{\varepsilon}$ is the average energy dissipation rate, and 
using dimensional analysis, {Kolmogorov and Hinze first  
	derived an estimate for} the maximum droplet size for which surface tension is able to resist the pressure fluctuations 
\cite{Kolmogorov1949,Hinze1955,Garrett2000}:\begin{equation}
d_H = \left(\frac{We_c}{2}\right)^{3/5}\left(\frac{\rho_c}{\sigma}\right)^{-3/5}\lra{\varepsilon}^{-2/5}.
\label{eq:hinze}
\end{equation}
This is referred to as the Hinze scale, and it is the only length scale that
can be obtained from $\rho_c, \sigma, \lra{\varepsilon}$.
For $d\gtrsim d_H$ surface tension forces cannot resist pressure fluctuations, 
and the droplets break. 
A fragmentation cascade is thus triggered, with a power-law size distribution  $\mathcal{N}(d)\sim d^{-10/3} $ obtained by dimensional analysis solely assuming locality~\cite{Garrett2000}.
Empirical evidences 
seem to confirm this distribution \cite{Garrett2000,Deane2002,blenkinsopp2010bubble,wang2016high,Deike:2016ir,Chan2020a,Deike2022}.
Yet, recent experiments on a single-bubble contradicts this physical picture \cite{qi2022fragmentation}, and
the definition of the critical Weber number $We_c$ remains ambiguous and somewhat heuristic,
with values in the literature spanning more than one order of magnitude~\cite{Deane2002,martinez1999breakup,Riviere2021}. 
Furthermore, the key feature of intermittency, that is the breaking of scale-invariance \cite{benzi1984multifractal,meneveau1987simple,Boffetta2008},  has not been considered in the analysis.

Even less clear is the behavior for $d\lesssim d_H$, where observations seem to indicate a different power-law spectrum $\mathcal{N}(d)\sim d^{-3/2} $ \cite{Deane2002,Deike2022}; laboratory experiments are difficult and no fundamental study of the collective turbulent dynamics of droplets or bubbles is currently available.

To overcome these difficulties and fully understand the problem by gaining access to quantities difficult to measure in the laboratory, 
we carry out a large campaign of direct numerical simulations (DNS) of turbulent multiphase flows.
The simulations are performed at an unprecedented high-level of multiphase turbulence, $Re_\lambda\approx137$, varying the surface tension,
volume fraction
$\alpha$, 
and the ratio of the two fluid viscosity.
Key to our understanding is the scale-by-scale energy budget, accessible only in numerical experiments (
See Material and Methods for the details on the theoretical tools and the setup employed).

\section*{
	Scale interpretation of droplet size spectrum}
\Cref{fig:dsd_render}(a) shows a visualisation of the mixture for a volume fraction of the dispersed phase $\alpha=0.1$ and viscosity ratio $1$. 
The dispersed phase, initially a single droplet, 
is organising over different scales and shapes, from large-scale drops to smaller filaments and even smaller droplets.
The droplet size distribution is shown for different volume fractions in the main panel of \Cref{fig:dsd_render} with the two scaling regimes introduced above clearly visible.
The small-scale range is similar for the different cases, whereas the large-scale distribution  is  sensitive to changes in the volume fraction. 
At large volume fractions, nevertheless, our results fully support the two empirically-proposed laws, remarkably neatly for $\alpha=0.5$.
For small
volume fractions of the dispersed phase, instead, the spectrum appears to fall exponentially at large scales.
Interestingly, the distribution spectra are only slightly altered when changing the viscosity,
as shown by the collapse of the data at fixed volume fraction  in panel (\textit{b}).

\begin{figure}[h!]
	{\includegraphics[width=0.48\columnwidth]{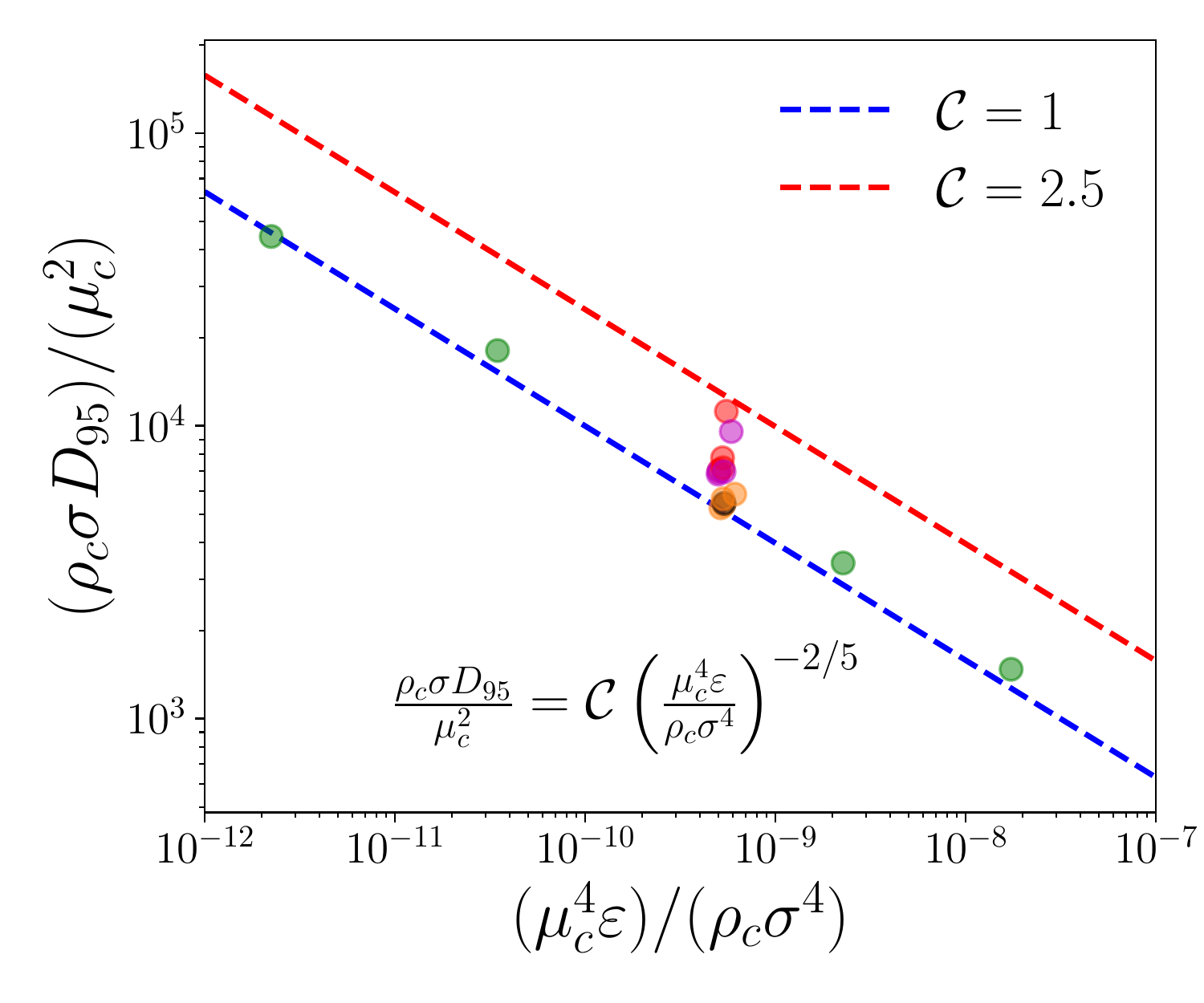}
		\put(-150,95){$We_\mathcal{L}$}
		\begin{tikzpicture}[remember picture,overlay]
		\node[coordinate] (A) at (-3,2.8) {};
		\node[coordinate] (B) at (-4.8,3.94) {};
		\draw[thick,->] (A) -- (B);
		\end{tikzpicture}	
		\begin{tikzpicture}[remember picture,overlay]
		\node[coordinate] (A) at (-2.9,2.5) {};
		\node[coordinate] (B) at (-2.9,4.2) {};
		\draw[thick,->] (A) -- (B);
		\end{tikzpicture}
		\put(-80,110){$\alpha,$ $\mu_d/\mu_c$}
		\put(-230,190){\large\textbf{(\textit{a})}}
	}
	
	\includegraphics[width=0.48\columnwidth]{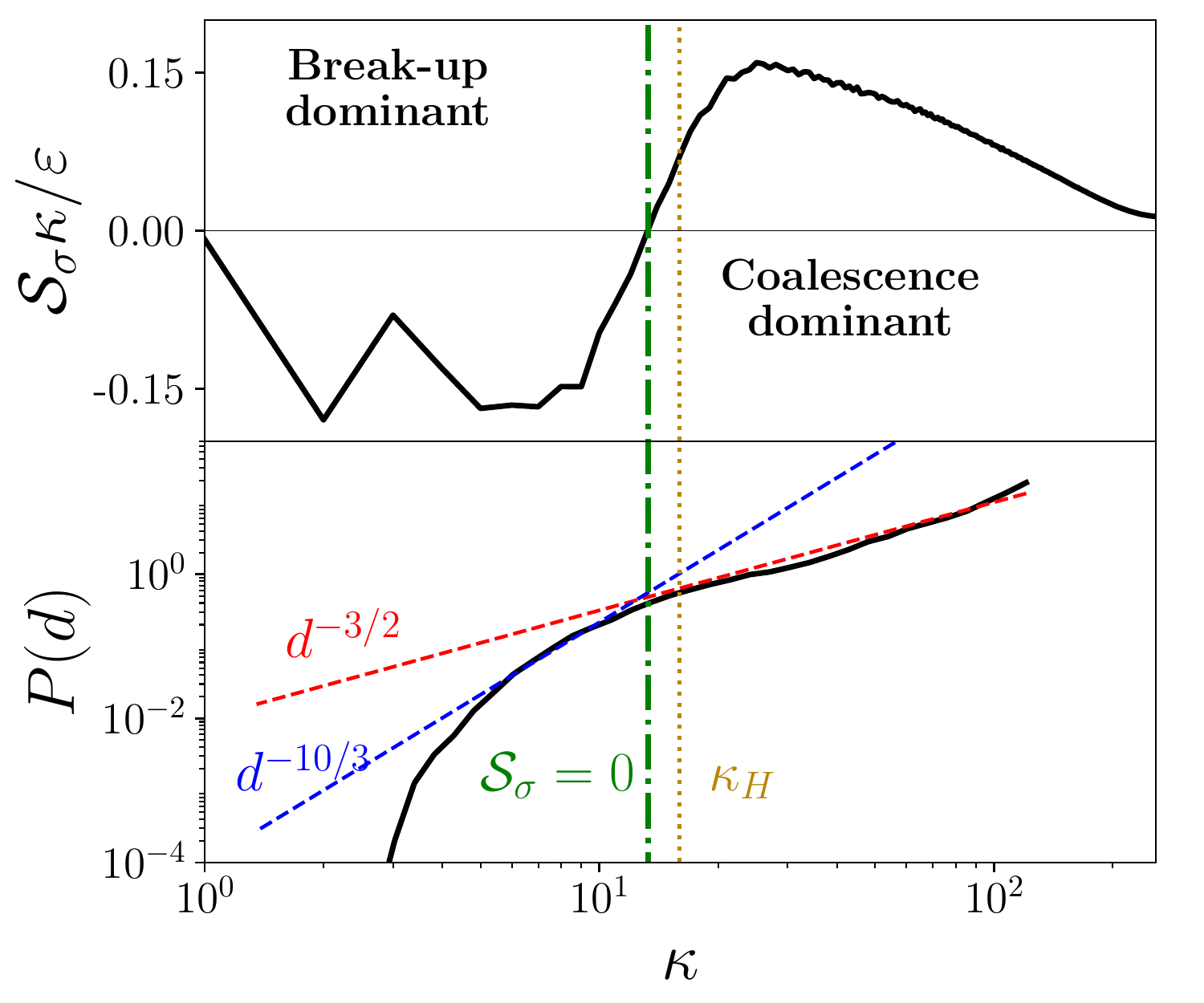}
	\put(-230,190){\large\textbf{(\textit{b})}}
	\caption{
		(a) Maximum droplet diameter $D_{95}$ as a function of the energy input $\varepsilon$, normalized as in \cite{Hinze1955}. We performed several simulations, varying one parameter (\emph{i.e.} $\alpha$, \mur and \wel) at the time. We show results for different $\alpha$ (\ref{fam_a}) at \wel$=42.6$ and \mur$=1$;
		different $We_{\mathcal{L}}$ (\ref{fam_we}) at \mur$=1$ and $\alpha=0.03$;
		different $\mu_d/\mu_c$  (\ref{fam_mu10}) at $\alpha=0.1$ and \wel$=42.6$,
		and different $\mu_d/\mu_c$  (\ref{fam_mu3}) at $\alpha=0.03$ and \wel$=42.6$.
		The arrows indicate data obtained with increasing values of each parameter.
		The blue line shows  $\mathcal{C}=\left(We_c/2\right)^{3/5}=1$, whereas $\mathcal{C}=0.765$ was originally proposed~\cite{Hinze1955}.
		(b) The relationship between surface tension energy transfer and droplet-size-distribution. Top panel shows the term $\mathcal{S}_\sigma$ from \Cref{eq:sbs}, normalized by the average energy dissipation and multiplied by $\kappa$ to increase visibility. The bottom panel shows the droplet-size-distribution $P(d)$, versus the wavenumber $\kappa=2\pi/d$. The vertical lines show $\mathcal{S}_{\sigma}=0, $ (green-dashed) and the Hinze scale wavenumber computed using $\mathcal{C}=1$ (gold-dotted).
	}
	\label{fig:hinzeFit}
\end{figure}

The cross-over scale separating the two power-laws is the Hinze scale, and is usually determined from the value of $We_c$.
In \Cref{fig:hinzeFit}(\textit{a}), we plot this parameter as a function of the two dimensionless groups  $\rho_c\sigma D_{95}/\mu_c^2$ and $\mu_c^4\varepsilon/\rho_c\sigma^4$, where $D_{95}$ is
the diameter for which $95\%$ of the total mass is enclosed in droplets smaller than $D_{95}$ \cite{Hinze1955}. 
Our results confirm that a unique value of the coefficient $\mathcal{C}=(We_c/2)^{3/5}$ cannot be used to fit all the data.
In particular, when surface tension is  varied while taking the viscosity of the two fluids equal, data nicely collapse on a line of slope $\mathcal{C} \approx1$.
However, increasing $\mu_d/\mu_c $ and $\alpha$, while maintaining 
the dissipation 
$\lra{\varepsilon}$ almost constant,  $\mathcal{C}$ should be increased to about 2.5 to fit the data.
Note that the range $\mathcal{C}\in[1,2.5]$ corresponds to a critical Weber number in the range  $2<We_c<9.5$; with even larger deviations observed in literature \cite{Riviere2021}.
This variability of $We_c$, 
limits the applicability of the original definition of the Hinze scale. 

We can however  provide  an accurate calculation of the cross-over scale by analysing the energy cascade as shown in the following. 
The scale-by-scale energy fluxes~\cite{Frisch1995a,Pope2009} 
written in Fourier space read (see Material and Methods):
\begin{equation}
\partial_t \mathcal{E}(\kappa) = T(\kappa) + \mathcal{D}(\kappa) + \mathcal{S}_\sigma(\kappa) +\mathcal{F}(\kappa)~,
\label{eq:sbs}
\end{equation}
where $\mathcal{E}(\kappa)$ is the  energy spectrum, $T(\kappa)$ the energy transfer due to the nonlinear term, $\mathcal{D}(\kappa)$ the viscous dissipation, 
$\mathcal{S}_\sigma(\kappa)$ the work of the surface tension force, and 
$\mathcal{F}(\kappa)$ is the power injected by the forcing used to maintain the turbulence. 
In \Cref{fig:hinzeFit}(\textit{b}), we display the net transfer due to the action of surface tension forces and the droplet size distribution versus the wavenumber $\kappa$.
The comparison of the scale-by-scale budget and the size-distribution remarkably points out that the cross-over scale is unambiguously defined as the length at which the work made by surface tension is zero, called hereafter as $d_{H\sigma}$.
It is found that this is generally different from the standard $d_H$ computed through (\ref{eq:hinze}).

At larger scales (small $\kappa$), 
$\mathcal{S}_\sigma(\kappa)<0$
which implies drainage by the surface tension forces. 
This indicates that, at these scales,
cohesive forces are not able to resist disruptive turbulent eddies and we should expect fragmentation to dominate.
In this regime, assuming statistical scale locality, that is droplets are broken only by eddies of comparable size, it is possible to obtain  the $-10/3$ power-law for the droplet-size-distribution \cite{Garrett2000}.
As in all turbulent cascades \cite{alexakis2018cascades}, the local picture cannot be strictly true.
Indeed, since droplets are stable due to surface tension when $d<d_{H\sigma}$, the local cascade would induce an accumulation of droplets at $\ell\approx d_{H\sigma}$ and no smaller ones, at least in a statistical sense.
We therefore expect this picture to be accurate only for  $d\gtrapprox d_{H\sigma}$,
and a non-local process to determine large-scale events. In fact, 
two similar daughter drops and some smaller droplets form as a result of the breakup of large droplets,  $d\gg d_{H\sigma}$,
as
confirmed experimentally \cite{Chan2020a,Riviere2021a}. It is worth emphasising that the cascade can be still approximated as local, since small droplets have negligible volume. Yet, non-local effects are necessary to explain the existence of droplets with $d < d_{H\sigma}$.

At small scales, droplets cannot break up as the surface tension is larger than the dynamic pressure, rather they should   coalesce 
for they try to minimize the free energy, that is the surface area.
In this case, we find positive work made by surface tension forces, that is an inverse cascade increasing the flow kinetic energy, as shown by the rightmost part of the spectrum in \Cref{fig:hinzeFit}(\textit{b}).
For geometric reasons, it is more likely to have {collision and} coalescence between small and large droplets, which promotes the non-locality of the inverse cascade.  
The introduction of a second large scale other than the droplet diameter allows us
to retrieve the  $-3/2$ scaling~\cite{Garrett2000,Deane2002}.
The closer the droplets are to the Hinze scale from below, \emph{i.e.} $d\lessapprox d_{H\sigma}$, the more local is the process, and  droplets of similar size are more likely to interact. For $d= d_{H\sigma}$ both mechanisms occur and the net energy transfer from surface tension is thus exactly zero 
\emph{i.e.} $\mathcal{S}_\sigma (\kappa_H)=0$, with  $\kappa_H=2\pi/d_H$.
At this scale, we observe the transition between the $-10/3$ and the $-3/2$ power-laws. 
The non-locality and randomness of the coalescence events suggest that this process is not likely the only source for positive surface tension work at small scales. 
In fact, random coalescence events should have a considerably high frequency to sustain small scale agitation, which is unlikely at very low volume fractions. 
{One possible production mechanism is the collision between a droplet and vortices of similar size but unable to break the interface, producing velocity fluctuations of scale smaller that the droplet size.}
A precise estimation of all the mechanisms leading to small-scale agitation is difficult to obtain from statistical data and requires additional ad-hoc numerical experiments. 
Our physical picture based on energy considerations are to be true in a statistical sense, but we expect 
it to be qualitatively true also for each single realisation.
Therefore, in order to gain a better understanding of the small-scale dynamics, we remove coalescence from the picture, and study the behavior in wave-number space of a single droplet break up.

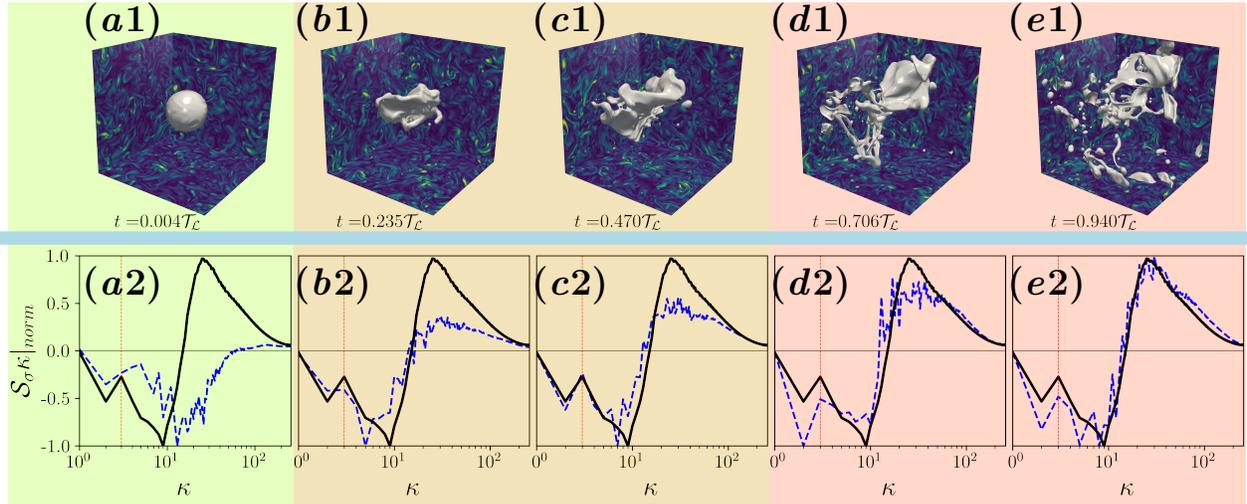
\begin{figure*}[h]
	\includegraphics[width=\textwidth]{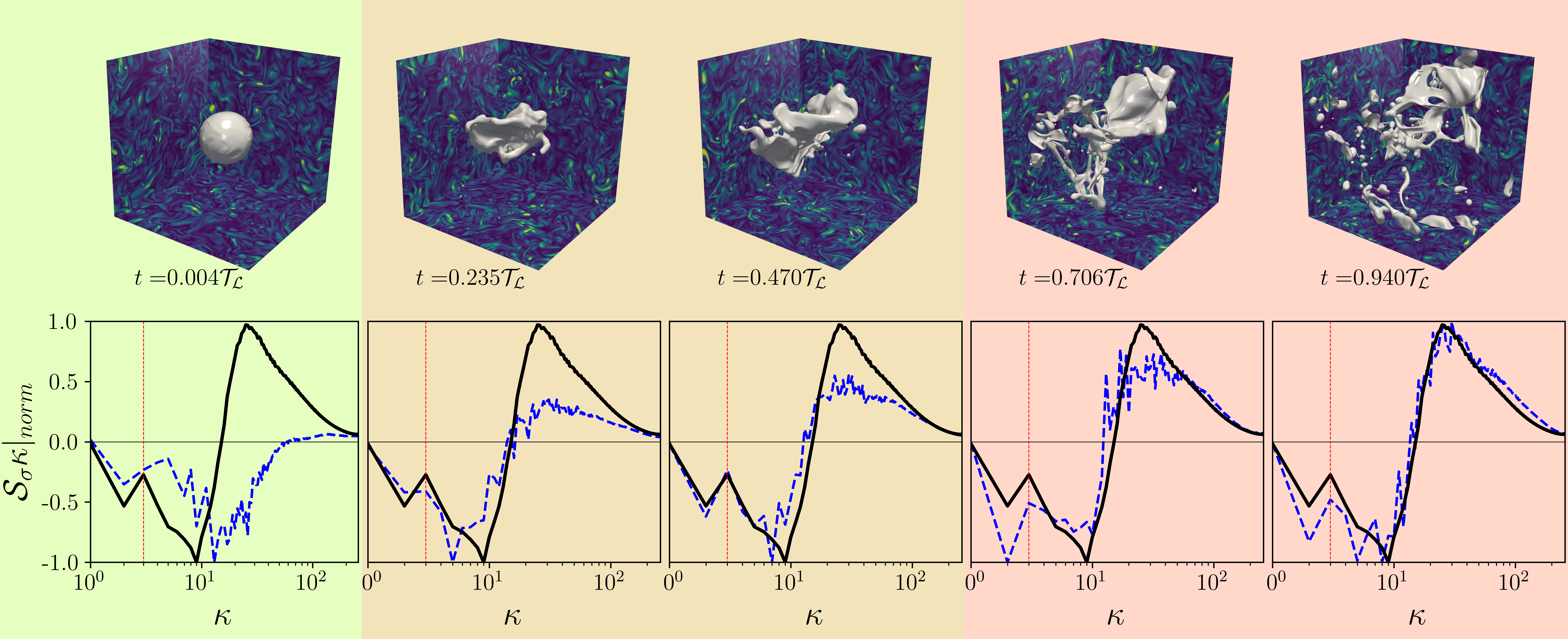}
	\put(-440,180){\large\textbf{(\textit{a}1)}}
	\put(-360,180){\large\textbf{(\textit{b}1)}}
	\put(-270,180){\large\textbf{(\textit{c}1)}}
	\put(-180,180){\large\textbf{(\textit{d}1)}}
	\put(-90,180){\large\textbf{(\textit{e}1)}}
	\put(-440,80){\large\textbf{(\textit{a}2)}}
	\put(-360,80){\large\textbf{(\textit{b}2)}}
	\put(-270,80){\large\textbf{(\textit{c}2)}}
	\put(-180,80){\large\textbf{(\textit{d}2)}}
	\put( -90,80){\large\textbf{(\textit{e}2)}}
	\begin{tikzpicture}[remember picture,overlay]
	\node[coordinate] (A) at (-18   ,3.57) {};
	\node[coordinate] (B) at (1     ,3.57) {};
	\draw[-Triangle Cap,line width=5pt, color=lightblue] (A) -- (B);
	\end{tikzpicture}
	\caption{Time-sequence of a single droplet breakup in turbulence. The simulation is performed at $\mu_d/\mu_c=1$ and $We_\mathcal{L}=42.6$. The volume fraction is set to $\alpha=0.0775$ (1 droplet of diameter d), so that $\kappa_d=2\pi/d\sim 3$. Top panels (\textit{a}1-\textit{e}1) show the temporal evolution of the interface during breakup, while vorticity is shown on projected planes. Bottom panels (\textit{a}2-\textit{e}2) show the surface tension energy transfer function $\mathcal{S}_\sigma$ (see \Cref{eq:sbs} and Material and Methods) normalized by its maximum values to improve readability. In each plot, in dashed-blue the instantaneous value of $\mathcal{S}_\sigma$ (corresponding to the snapshot above), while the time-averaged value in statistical-stationary condition is reported using a black line. The dotted red line indicates $\kappa_d$ and time $t$ is normalized with the large-eddies turnover time $\mathcal{T}_\mathcal{L}$.
	}
	\label{fig:strip}
\end{figure*}
\section*{Single-droplet breakup
}
The statistical interpretation of the break-up/coalescence process described so far relies on the assumption that a droplet can be considered as a spherical object,
such that local variations of the surface curvature are statistically negligible. 
Of course, this cannot be true for a single break-up, and one must provide a link between the single-droplet dynamics and the multiphase flow statistics.
The
morphological 
analysis of a droplet break-up in turbulence enables us to identify three stages, see \Cref{fig:strip}: \textit{incipient} deformation (green panels), \textit{sub-critical} deformation (yellow panels) and \textit{super-critical} deformation (red panels).
In the \textit{incipient} deformation stage, the droplet, originally spherical, deforms due to the interaction with the turbulence, panel (\textit{a}1). At this stage, the turbulent kinetic energy is mainly absorbed at large and intermediate scales, as shown by the work of the surface tension $\mathcal{S}_\sigma$ in panel (\textit{a}2). 
Deformation increases with time, progressively forming regions with high values of the curvature $\xi$ (panels \textit{b}1,\textit{c}1). At this \textit{sub-critical} stage,
work against surface tension is acting to deform larger interfaces while

smaller structures are produced
for which
interfacial forces are greater than turbulent pressure fluctuations. At these small scales, the interface tends to relax to a spherical shape,  
releasing energy to the surrounding flow (see areas of $\mathcal{S}_\sigma>0$ in panels \textit{b}2,\textit{c}2). 
It can be demonstrated that high local values of $\xi$ are directly responsible for high values of $\mathcal{S}_\sigma$ (see Material and Methods).
When \textit{super-critical} deformation is reached, the interface breaks and small droplets form, see panels (\textit{d}1,\textit{e}1). Energy is still absorbed for the deformation of large intefaces, 
while the 
coalescence of the small droplets minimizes the surface area and 
adds energy to the flow at small scales (panels \textit{d}1,\textit{e}1). 
Simulations pleasantly confirm that most of the fragmentation process is local, while sub-Hinze droplets are formed through a non-local dynamics.
Droplets at $d<d_H$ are thus characterised by an oscillatory motion, resulting from
deformations by
small vortices and viscosity and surface-tension driven 
relaxation.
Interestingly, droplet relaxation and coalescence have a similar energy footprint on the flow, as they both add energy at scales $\ell<d$. At the same time, coalescence contributes to the formation of larger drops, thus affecting energy transfer to larger scales, proving to be the source of non-locality from small to large scales. 

After a full large-eddy turnover time $\mathcal{T}_\mathcal{L}$, the instantaneous energy transfer due to surface tension forces, $\mathcal{S}_\sigma$, approaches the behavior at the statistically stationary state, corroborating the statistical picture obtained when many droplets are considered.
This shows that droplet dynamics
for sizes $d<d_{H\sigma}$ and local interface deformations at scale $\ell<d_{H\sigma}$ 
are associated to the transfer of energy to the carrier phase by interfacial forces.
Our simulations clearly show that energy spectra of the carrying phase are modulated by the droplets in a way fully consistent with the above physical picture (see Supporting Information).

\begin{figure}[h!]
	\includegraphics[width=0.48\columnwidth]{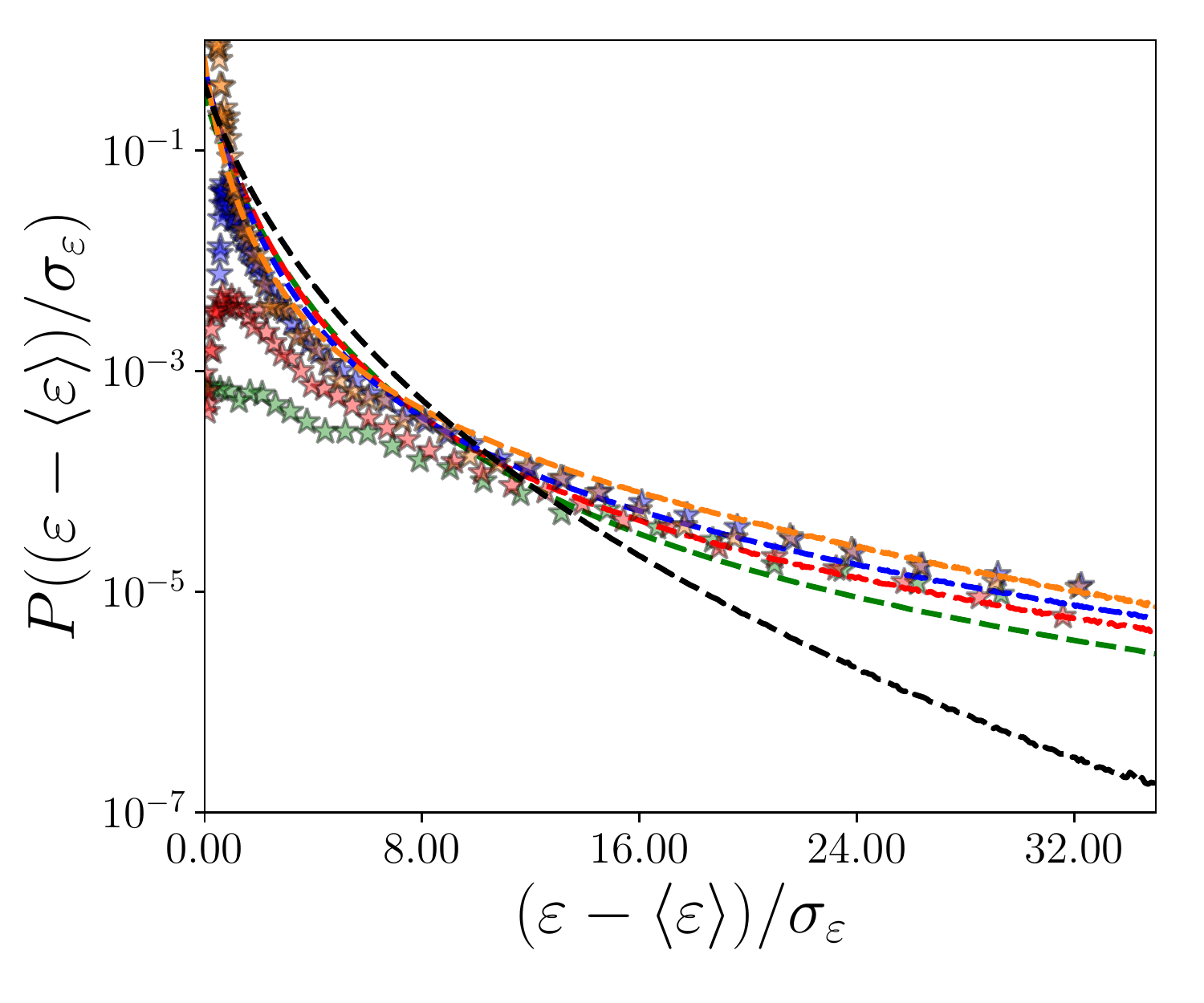}
	\put(-220,200){\large\textbf{(\textit{a})}}
	\put(-115,110){\includegraphics[scale=0.4]{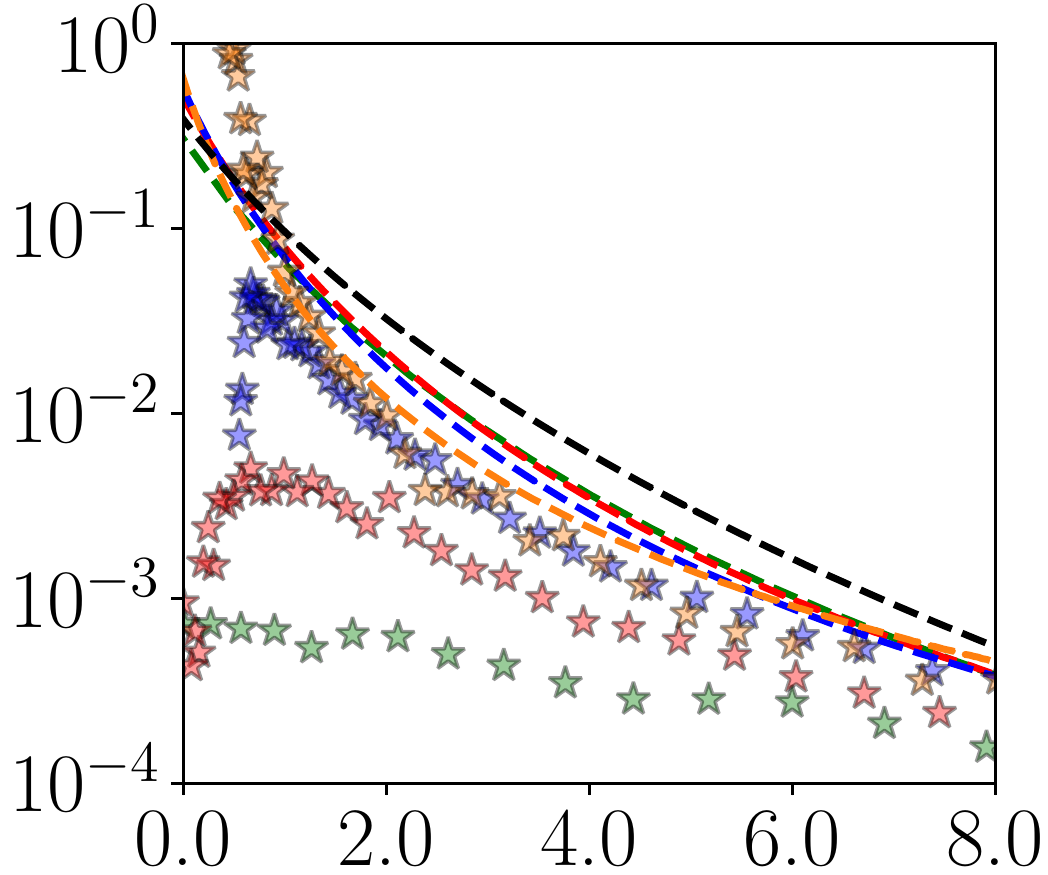}}\\
	\includegraphics[width=0.48\columnwidth]{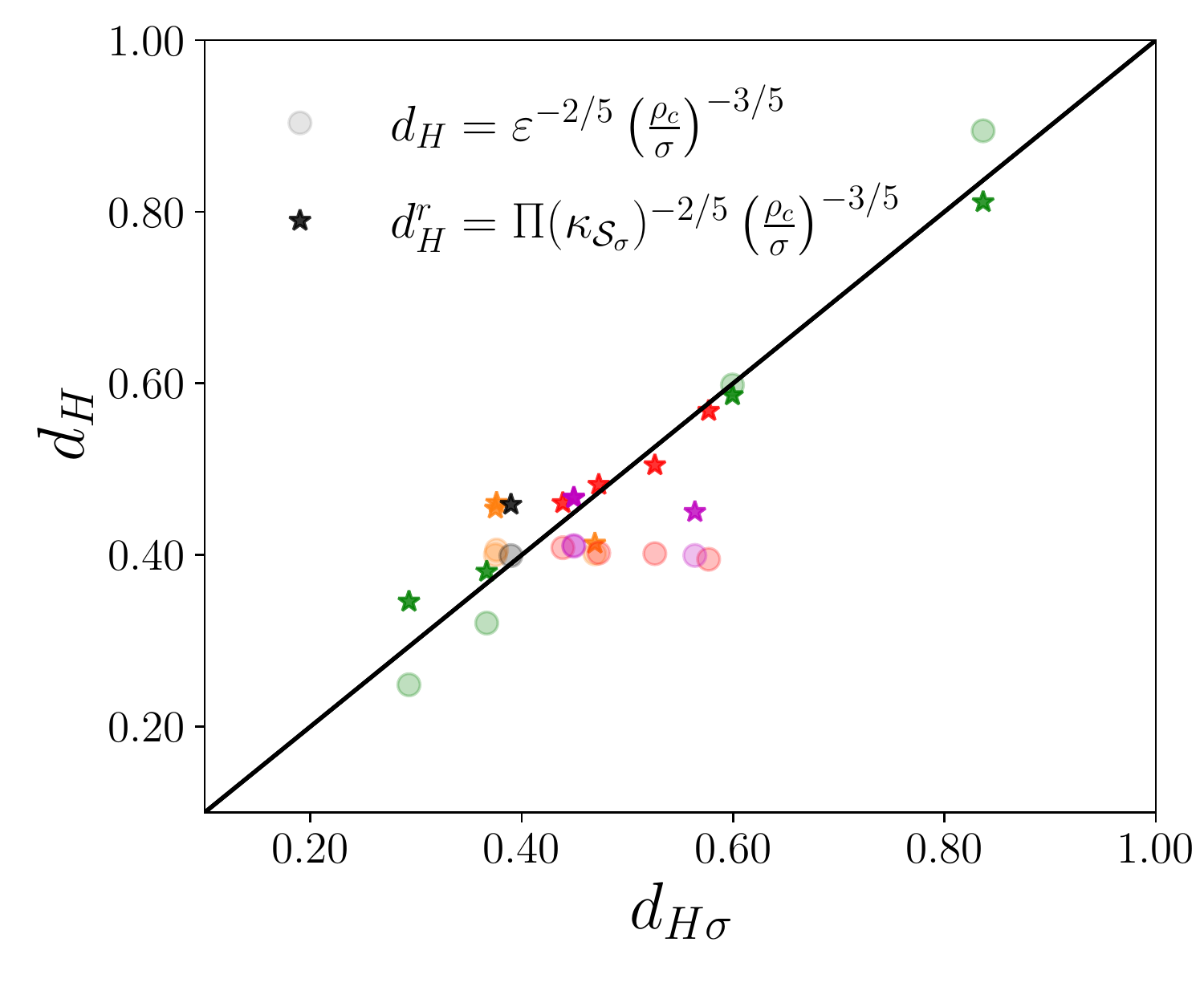}
	\put(-220,200){\large\textbf{(\textit{b})}}
	\caption{
		(\textit{a}) Probability-density-function for the normalized energy dissipation rate $\varepsilon$. Colored dashed lines show data from multiphase simulations (see legend in caption \Cref{fig:dsd_render}), while the single-phase data are shown in black. The stars indicate the values of epsilon computed from the DSD where $\varepsilon\propto d^{-5/2}$ and $P(\varepsilon)\propto d^{13/2}N(d)$. The inset shows the details for the PDF at low $\varepsilon$.
		(b) Comparison among the different methods used to compute the Hinze scale, \emph{i.e.} the original formulation $d_H$, and the proposed approach $d_{H_\sigma}$, obtained as the scale at which $\mathcal{S}_\sigma=0$. For $d_H$, 
		we show two formulations, namely the original formulation proposed in \cite{Hinze1955} (circles), and the novel interpretation $d_H^r$ (stars). The computation of $d_H^r$ uses the wavenumber-local non-linear fluxes at $\Pi(\kappa_{\mathcal{S}_\sigma})$, with $\kappa_{\mathcal{S}_\sigma}=2\pi/d_{H_\sigma}$, where a prefactor $0.8$ has been applied. A black diagonal line shows where $d_H^r=d_{H_\sigma}$, highlighting the improvement provided by the novel interpretation of the Hinze scale.   
		Colors are consistent with the previously used, see \Cref{fig:hinzeFit}
	}
	\label{fig:pdfeps}
\end{figure}

\section*{Intermittency}
\label{sec:interm}
The argument leading to the definition of the Hinze scale is based on 
mean turbulence properties,
yet turbulence is characterised by fluctuations exhibiting large deviations from  the mean values, \emph{i.e.}\ intermittency~\cite{Frisch1995a}.
The definition of a Hinze scale based on the energy fluxes takes implicitly  into account intermittency in an average sense, but it does not give any insight on the role of the extreme events on the droplet size distribution.
If
the turbulence determines the droplet size distribution
and the droplets modulate the turbulence,
it should be possible to relate the droplet size
distribution to the intermittency at the different scales.
If this is the case, we would be able to extract information about the turbulence modulation and dissipation rates
from the distribution size spectrum, more accessible to
experimental measurements.

Using \Cref{eq:hinze}, the probability distributions of $d$ and $\varepsilon$ (here intended as the space-local and instantaneous value of dissipation) may be related by $\varepsilon \sim d^{-5/2}$, so to obtain~\cite{Garrett2000}   
\begin{equation}
P(\varepsilon)\sim d^{13/2}\mathcal{N}(d)~.
\label{eq:prob}
\end{equation}
This relation implies that small droplets are correlated to high dissipation events (large $\varepsilon$), while large droplets are related to regions of small $\varepsilon$. 
This is beautifully confirmed 
in \Cref{fig:pdfeps}, where we compare the distribution $P(\varepsilon)$, computed according to \Cref{eq:prob},  with the dissipation distribution extracted  from the simulations.
The tail of the distribution,  \emph{i.e.}\ high $\varepsilon$, which corresponds to small scales and the range of diameters for which the droplet distribution scales as $d^{-3/2}$, is accurately described by \Cref{eq:prob} for all cases.
For small $\varepsilon$, \emph{i.e.}\ large $d$ (see inset), \Cref{eq:prob} is in general less accurate, yet improving with $\alpha$  (the whole dissipation distribution is reasonably well predicted for  $\alpha=0.5$). 
Eq.\ (\ref{eq:prob}) is expected to be true only in the inertial range, where the local theory of turbulence is valid, and 
indeed for increasing values of $\alpha$ more droplets are found in the inertial range (see Supporting Information). 
Remarkably, \Cref{fig:pdfeps} also shows that the degree of intermittency  is much higher in multiphase flows than in single-phase turbulence, indicating an increase of extreme events at small scales when coalescence 
is also important. 
This is thought to be due to the vorticity creation related to interfaces, which seems to be a crucial feature of intermittency for all practical applications \cite{buaria2022vorticity}.

We are now in the position to provide a statistical description which relates the intermittent dissipation to the Hinze scale.
The rationale underlying \Cref{eq:hinze} is the competition between capillary forces and the turbulent stresses, which can be related to the turbulent dissipation rate. 
As shown by the refined-Kolmogorov theory \cite{Kolmogorov1962,Frisch1995a,Boffetta2008,Dubrulle2019}, high-energy intermittent events are occurring within the dissipation range, \emph{i.e.} localized at small scales \cite{Buzzicotti2020}. The correlation (not necessarily causality) 
between small droplets and high values of $\varepsilon$, typical of small scales, suggests that a reinterpretation of the breakup theory could be obtained by  defining a local random variable $\varepsilon_\ell$ as the integral over a sphere of diameter $\ell$ of the dissipation field $\varepsilon({\bf x},t)$.
To include intermittency in the picture, we relate the
turbulent forces to the local dissipation rate $\varepsilon_\ell$ at the relevant scale $\ell$;
for our purpose it is safe to assume valid the Kolmogorov refined-similarity hypothesis~\cite{Kolmogorov1962}, which states that $(\varepsilon_\ell \ell)^{1/3}$ is statistically equivalent to $u_\ell$. 
In this framework, the local dissipation is  estimated in term of the cascade flux as $\varepsilon_\ell \sim u_\ell^3/\ell \sim \Pi_\ell$.
Hence, if $\kappa_{\mathcal{S}_\sigma}$ is the wavenumber corresponding to $\mathcal{S}_\sigma=0$ in the shell-by-shell energy budget,  the energy flux $\Pi(\kappa_{\mathcal{S}_\sigma})=\sum_{0}^{\kappa_{\mathcal{S}_\sigma}}T(\kappa)$ can be associated with the scale-local velocity fluctuations at the Hinze scale. 
By replacing $\varepsilon$ with $\Pi(\kappa_{\mathcal{S}_\sigma})$ in (\ref{eq:hinze}) we obtain a refined definition  of the Hinze scale:
\begin{equation}
d_H^r = \left(\frac{\rho_c}{\sigma}\right)^{-3/5}\Pi(\kappa_{\mathcal{S}_\sigma})^{-2/5}.
\label{eq:localHinze}
\end{equation}
Note that our modified picture reduces to the original Hinze prediction at low volume fractions when turbulence modulation is negligible, consistently with the hypothesis discussed in \cite{Hinze1955}, as $\Pi_\ell\sim\varepsilon_\ell\sim\langle\varepsilon\rangle$ and the surface tension energy flux is vanishing.
The prediction obtained from this relation are compared to classic KH theory in 
\Cref{fig:pdfeps}. Most of the values of $d_H^r$ from Eq.\ (\ref{eq:localHinze}) coincide with the values of $d_{H_\sigma}= 2\pi/\kappa_{\mathcal{S}_\sigma}$ when using a pre-factor 0.8, clearly improving over the standard definition in Eq.\ \ref{eq:hinze}.
The pre-factor might be related to finite-$Re_\lambda$ effects.

\section*{Discussion}
In the present study, we provide evidences for two crucial hypothesis on the dynamics of multiphase flows when turbulence modulation by the dispersed phase cannot be neglected.
First, we propose an unambiguous definition of the Hinze scale based on the analysis of the scale-by-scale energy transfer: this is the scale where the net energy transferred by the interfacial forces is zero. This scale separates two regimes:
the dynamics at large scales ($d>d_H$) is characterised by a local fragmentation cascade
and a net loss of energy of the larger flow structures when interacting with the dispersed phase. 
Droplet coalescence and interfacial deformations, instead,
dominate at smaller scales where energy is re-injected as a result of a non-local process, further extending the dissipative range towards smaller scales \cite{crialesi2022}
Secondly, we demonstrate the link between the droplet size distribution, with pivoting Hinze scale, and the turbulent intermittency. Intermittent rare events at small scale increase in the presence of droplets, with a probability  proportional to  $d^{13/2}\mathcal{N}(d)$. 
In addition, we show that a consistent new  definition of the Hinze scale can be achieved considering  local intermittent fluctuations of the dissipation in the spirit of the Kolmogorov refined similarity hypothesis.

Although energy fluxes are difficult to measure, the scale $d_H^r=\left(\rho_c/\sigma\right)^{-3/5}\Pi(\kappa_{\mathcal{S}_\sigma})^{-2/5}$
does not depend on any fitting parameter.
As $d_H^r\approx d_{H\sigma}$, this scale can be estimated from measurements of the droplet-size spectrum and it can be used to estimate the local energy flux $\Pi(\kappa_{\mathcal{S}_\sigma})$.
Knowing  $\langle\varepsilon\rangle$, one can obtain $\Pi^\sigma(\kappa_{\mathcal{S}_\sigma}) = \sum_{0}^{\kappa_{\mathcal{S}_\sigma}} S_\sigma=\langle\varepsilon\rangle - \Pi(\kappa_{\mathcal{S}_\sigma})$, which is the energy net flux across the wavenumber $\kappa_{\mathcal{S}_\sigma}$  due to surface tension forces, \emph{i.e.} the maximum value of $\Pi^\sigma$ (see \Cref{fig:hinzeFit}). This computation provides a direct quantification of turbulence modulation due to interfacial forces in multiphase flows. 
Finally, we stress that, based on our results, from the sole observation of the droplet/bubble size-distribution, one could infer $\langle\varepsilon\rangle$ and the dissipation fluctuations (see \Cref{fig:pdfeps}), hence the most relevant features of the flow.

Dynamical observation of the breaking of a single droplet nicely confirms the statistical picture from a pure geometrical/energetic point of view.
We have thus demonstrated that 
a droplet of size $d=2\pi/\kappa$ influences the energy transfer at $\kappa$ through its next topological transformation. In particular, for sub-Hinze inclusions, $d<d_H^r$, we find energy injection associated to creation of  small-scale vorticity, whereas we document energy absorption for super-Hinze droplets $d>d_H^r$, bound to break up. 
The present results provide insights for future coarse-graining modelling of droplet/bubbles dynamics.

\bibliography{references_new.bib}

\begin{thebibliography}{47}%
\makeatletter
\providecommand \@ifxundefined [1]{%
 \@ifx{#1\undefined}
}%
\providecommand \@ifnum [1]{%
 \ifnum #1\expandafter \@firstoftwo
 \else \expandafter \@secondoftwo
 \fi
}%
\providecommand \@ifx [1]{%
 \ifx #1\expandafter \@firstoftwo
 \else \expandafter \@secondoftwo
 \fi
}%
\providecommand \natexlab [1]{#1}%
\providecommand \enquote  [1]{``#1''}%
\providecommand \bibnamefont  [1]{#1}%
\providecommand \bibfnamefont [1]{#1}%
\providecommand \citenamefont [1]{#1}%
\providecommand \href@noop [0]{\@secondoftwo}%
\providecommand \href [0]{\begingroup \@sanitize@url \@href}%
\providecommand \@href[1]{\@@startlink{#1}\@@href}%
\providecommand \@@href[1]{\endgroup#1\@@endlink}%
\providecommand \@sanitize@url [0]{\catcode `\\12\catcode `\$12\catcode
  `\&12\catcode `\#12\catcode `\^12\catcode `\_12\catcode `\%12\relax}%
\providecommand \@@startlink[1]{}%
\providecommand \@@endlink[0]{}%
\providecommand \url  [0]{\begingroup\@sanitize@url \@url }%
\providecommand \@url [1]{\endgroup\@href {#1}{\urlprefix }}%
\providecommand \urlprefix  [0]{URL }%
\providecommand \Eprint [0]{\href }%
\providecommand \doibase [0]{https://doi.org/}%
\providecommand \selectlanguage [0]{\@gobble}%
\providecommand \bibinfo  [0]{\@secondoftwo}%
\providecommand \bibfield  [0]{\@secondoftwo}%
\providecommand \translation [1]{[#1]}%
\providecommand \BibitemOpen [0]{}%
\providecommand \bibitemStop [0]{}%
\providecommand \bibitemNoStop [0]{.\EOS\space}%
\providecommand \EOS [0]{\spacefactor3000\relax}%
\providecommand \BibitemShut  [1]{\csname bibitem#1\endcsname}%
\let\auto@bib@innerbib\@empty
\bibitem [{\citenamefont {Villermaux}\ and\ \citenamefont
  {Bossa}(2009)}]{Villermaux2009}%
  \BibitemOpen
  \bibfield  {author} {\bibinfo {author} {\bibfnamefont {E.}~\bibnamefont
  {Villermaux}}\ and\ \bibinfo {author} {\bibfnamefont {B.}~\bibnamefont
  {Bossa}},\ }\bibfield  {title} {\bibinfo {title} {{Single-drop fragmentation
  determines size distribution of raindrops}},\ }\href
  {https://doi.org/10.1038/nphys1340} {\bibfield  {journal} {\bibinfo
  {journal} {Nature Physics}\ }\textbf {\bibinfo {volume} {5}},\ \bibinfo
  {pages} {697} (\bibinfo {year} {2009})}\BibitemShut {NoStop}%
\bibitem [{\citenamefont {Keshavarz}\ \emph {et~al.}(2016)\citenamefont
  {Keshavarz}, \citenamefont {Houze}, \citenamefont {Moore}, \citenamefont
  {Koerner},\ and\ \citenamefont {McKinley}}]{Keshavarz2016}%
  \BibitemOpen
  \bibfield  {author} {\bibinfo {author} {\bibfnamefont {B.}~\bibnamefont
  {Keshavarz}}, \bibinfo {author} {\bibfnamefont {E.~C.}\ \bibnamefont
  {Houze}}, \bibinfo {author} {\bibfnamefont {J.~R.}\ \bibnamefont {Moore}},
  \bibinfo {author} {\bibfnamefont {M.~R.}\ \bibnamefont {Koerner}},\ and\
  \bibinfo {author} {\bibfnamefont {G.~H.}\ \bibnamefont {McKinley}},\
  }\bibfield  {title} {\bibinfo {title} {{Ligament Mediated Fragmentation of
  Viscoelastic Liquids}},\ }\href
  {https://doi.org/10.1103/PhysRevLett.117.154502} {\bibfield  {journal}
  {\bibinfo  {journal} {Physical Review Letters}\ }\textbf {\bibinfo {volume}
  {117}},\ \bibinfo {pages} {154502} (\bibinfo {year} {2016})}\BibitemShut
  {NoStop}%
\bibitem [{\citenamefont {Villermaux}(2020)}]{villermaux2020fragmentation}%
  \BibitemOpen
  \bibfield  {author} {\bibinfo {author} {\bibfnamefont {E.}~\bibnamefont
  {Villermaux}},\ }\bibfield  {title} {\bibinfo {title} {Fragmentation versus
  cohesion},\ }\href@noop {} {\bibfield  {journal} {\bibinfo  {journal}
  {Journal of Fluid Mechanics}\ }\textbf {\bibinfo {volume} {898}} (\bibinfo
  {year} {2020})}\BibitemShut {NoStop}%
\bibitem [{\citenamefont {Perlekar}\ \emph {et~al.}(2014)\citenamefont
  {Perlekar}, \citenamefont {Benzi}, \citenamefont {Clercx}, \citenamefont
  {Nelson},\ and\ \citenamefont {Toschi}}]{Perlekar2014}%
  \BibitemOpen
  \bibfield  {author} {\bibinfo {author} {\bibfnamefont {P.}~\bibnamefont
  {Perlekar}}, \bibinfo {author} {\bibfnamefont {R.}~\bibnamefont {Benzi}},
  \bibinfo {author} {\bibfnamefont {H.~J.}\ \bibnamefont {Clercx}}, \bibinfo
  {author} {\bibfnamefont {D.~R.}\ \bibnamefont {Nelson}},\ and\ \bibinfo
  {author} {\bibfnamefont {F.}~\bibnamefont {Toschi}},\ }\bibfield  {title}
  {\bibinfo {title} {{Spinodal decomposition in homogeneous and isotropic
  turbulence}},\ }\href {https://doi.org/10.1103/PhysRevLett.112.014502}
  {\bibfield  {journal} {\bibinfo  {journal} {Physical Review Letters}\
  }\textbf {\bibinfo {volume} {112}},\ \bibinfo {pages} {1} (\bibinfo {year}
  {2014})}\BibitemShut {NoStop}%
\bibitem [{\citenamefont {Girotto}\ \emph {et~al.}(2021)\citenamefont
  {Girotto}, \citenamefont {Benzi}, \citenamefont {Di~Staso}, \citenamefont
  {Scagliarini}, \citenamefont {Schifano},\ and\ \citenamefont
  {Toschi}}]{girotto2021}%
  \BibitemOpen
  \bibfield  {author} {\bibinfo {author} {\bibfnamefont {I.}~\bibnamefont
  {Girotto}}, \bibinfo {author} {\bibfnamefont {R.}~\bibnamefont {Benzi}},
  \bibinfo {author} {\bibfnamefont {G.}~\bibnamefont {Di~Staso}}, \bibinfo
  {author} {\bibfnamefont {A.}~\bibnamefont {Scagliarini}}, \bibinfo {author}
  {\bibfnamefont {S.~F.}\ \bibnamefont {Schifano}},\ and\ \bibinfo {author}
  {\bibfnamefont {F.}~\bibnamefont {Toschi}},\ }\bibfield  {title} {\bibinfo
  {title} {Build up of yield stress fluids via chaotic emulsification},\
  }\href@noop {} {\bibfield  {journal} {\bibinfo  {journal} {arXiv preprint
  arXiv:2111.12453}\ } (\bibinfo {year} {2021})}\BibitemShut {NoStop}%
\bibitem [{\citenamefont {Bakhuis}\ \emph {et~al.}(2021)\citenamefont
  {Bakhuis}, \citenamefont {Ezeta}, \citenamefont {Bullee}, \citenamefont
  {Marin}, \citenamefont {Lohse}, \citenamefont {Sun},\ and\ \citenamefont
  {Huisman}}]{Bakhuis2021}%
  \BibitemOpen
  \bibfield  {author} {\bibinfo {author} {\bibfnamefont {D.}~\bibnamefont
  {Bakhuis}}, \bibinfo {author} {\bibfnamefont {R.}~\bibnamefont {Ezeta}},
  \bibinfo {author} {\bibfnamefont {P.~A.}\ \bibnamefont {Bullee}}, \bibinfo
  {author} {\bibfnamefont {A.}~\bibnamefont {Marin}}, \bibinfo {author}
  {\bibfnamefont {D.}~\bibnamefont {Lohse}}, \bibinfo {author} {\bibfnamefont
  {C.}~\bibnamefont {Sun}},\ and\ \bibinfo {author} {\bibfnamefont {S.~G.}\
  \bibnamefont {Huisman}},\ }\bibfield  {title} {\bibinfo {title}
  {{Catastrophic Phase Inversion in High-Reynolds-Number Turbulent
  Taylor-Couette Flow}},\ }\href
  {https://doi.org/10.1103/PhysRevLett.126.064501} {\bibfield  {journal}
  {\bibinfo  {journal} {Physical Review Letters}\ }\textbf {\bibinfo {volume}
  {126}},\ \bibinfo {pages} {64501} (\bibinfo {year} {2021})},\ \Eprint
  {https://arxiv.org/abs/2010.03200} {arXiv:2010.03200} \BibitemShut {NoStop}%
\bibitem [{\citenamefont {Garrett}\ \emph {et~al.}(2000)\citenamefont
  {Garrett}, \citenamefont {Li},\ and\ \citenamefont {Farmer}}]{Garrett2000}%
  \BibitemOpen
  \bibfield  {author} {\bibinfo {author} {\bibfnamefont {C.}~\bibnamefont
  {Garrett}}, \bibinfo {author} {\bibfnamefont {M.}~\bibnamefont {Li}},\ and\
  \bibinfo {author} {\bibfnamefont {D.}~\bibnamefont {Farmer}},\ }\bibfield
  {title} {\bibinfo {title} {{The connection between bubble size spectra and
  energy dissipation rates in the upper ocean}},\ }\href
  {https://doi.org/10.1175/1520-0485(2000)030<2163:TCBBSS>2.0.CO;2} {\bibfield
  {journal} {\bibinfo  {journal} {Journal of Physical Oceanography}\ }\textbf
  {\bibinfo {volume} {30}},\ \bibinfo {pages} {2163} (\bibinfo {year}
  {2000})}\BibitemShut {NoStop}%
\bibitem [{\citenamefont {Gao}\ \emph {et~al.}(2021)\citenamefont {Gao},
  \citenamefont {Deane},\ and\ \citenamefont {Shen}}]{Gao2021}%
  \BibitemOpen
  \bibfield  {author} {\bibinfo {author} {\bibfnamefont {Q.}~\bibnamefont
  {Gao}}, \bibinfo {author} {\bibfnamefont {G.~B.}\ \bibnamefont {Deane}},\
  and\ \bibinfo {author} {\bibfnamefont {L.}~\bibnamefont {Shen}},\ }\bibfield
  {title} {\bibinfo {title} {{Bubble production by air filament and cavity
  breakup in plunging breaking wave crests}},\ }\href
  {https://doi.org/10.1017/jfm.2021.890} {\bibfield  {journal} {\bibinfo
  {journal} {Journal of Fluid Mechanics}\ }\textbf {\bibinfo {volume} {929}},\
  \bibinfo {pages} {A44} (\bibinfo {year} {2021})}\BibitemShut {NoStop}%
\bibitem [{\citenamefont {Deike}(2022)}]{Deike2022}%
  \BibitemOpen
  \bibfield  {author} {\bibinfo {author} {\bibfnamefont {L.}~\bibnamefont
  {Deike}},\ }\bibfield  {title} {\bibinfo {title} {{Mass Transfer at the
  Ocean–Atmosphere Interface: The Role of Wave Breaking, Droplets, and
  Bubbles}},\ }\href {https://doi.org/10.1146/annurev-fluid-030121-014132}
  {\bibfield  {journal} {\bibinfo  {journal} {Annual Review of Fluid
  Mechanics}\ }\textbf {\bibinfo {volume} {54}},\ \bibinfo {pages} {191}
  (\bibinfo {year} {2022})}\BibitemShut {NoStop}%
\bibitem [{\citenamefont {Berny}\ \emph {et~al.}(2021)\citenamefont {Berny},
  \citenamefont {Popinet}, \citenamefont {S{\'{e}}on},\ and\ \citenamefont
  {Deike}}]{Berny2021}%
  \BibitemOpen
  \bibfield  {author} {\bibinfo {author} {\bibfnamefont {A.}~\bibnamefont
  {Berny}}, \bibinfo {author} {\bibfnamefont {S.}~\bibnamefont {Popinet}},
  \bibinfo {author} {\bibfnamefont {T.}~\bibnamefont {S{\'{e}}on}},\ and\
  \bibinfo {author} {\bibfnamefont {L.}~\bibnamefont {Deike}},\ }\bibfield
  {title} {\bibinfo {title} {{Statistics of Jet Drop Production}},\ }\href
  {https://doi.org/10.1029/2021GL092919} {\bibfield  {journal} {\bibinfo
  {journal} {Geophysical Research Letters}\ }\textbf {\bibinfo {volume} {48}},\
  \bibinfo {pages} {1} (\bibinfo {year} {2021})}\BibitemShut {NoStop}%
\bibitem [{\citenamefont {Jiang}\ \emph {et~al.}(2022)\citenamefont {Jiang},
  \citenamefont {Rotily}, \citenamefont {Villermaux},\ and\ \citenamefont
  {Wang}}]{Jiang2022}%
  \BibitemOpen
  \bibfield  {author} {\bibinfo {author} {\bibfnamefont {X.}~\bibnamefont
  {Jiang}}, \bibinfo {author} {\bibfnamefont {L.}~\bibnamefont {Rotily}},
  \bibinfo {author} {\bibfnamefont {E.}~\bibnamefont {Villermaux}},\ and\
  \bibinfo {author} {\bibfnamefont {X.}~\bibnamefont {Wang}},\ }\bibfield
  {title} {\bibinfo {title} {{Submicron drops from flapping bursting
  bubbles}},\ }\href {https://doi.org/10.1073/pnas.2112924119} {\bibfield
  {journal} {\bibinfo  {journal} {Proceedings of the National Academy of
  Sciences}\ }\textbf {\bibinfo {volume} {119}},\ \bibinfo {pages}
  {e2112924119} (\bibinfo {year} {2022})}\BibitemShut {NoStop}%
\bibitem [{\citenamefont {Skartlien}\ \emph {et~al.}(2013)\citenamefont
  {Skartlien}, \citenamefont {Sollum},\ and\ \citenamefont
  {Schumann}}]{Skartlien2013}%
  \BibitemOpen
  \bibfield  {author} {\bibinfo {author} {\bibfnamefont {R.}~\bibnamefont
  {Skartlien}}, \bibinfo {author} {\bibfnamefont {E.}~\bibnamefont {Sollum}},\
  and\ \bibinfo {author} {\bibfnamefont {H.}~\bibnamefont {Schumann}},\
  }\bibfield  {title} {\bibinfo {title} {{Droplet size distributions in
  turbulent emulsions: Breakup criteria and surfactant effects from direct
  numerical simulations}},\ }\bibfield  {journal} {\bibinfo  {journal} {Journal
  of Chemical Physics}\ }\textbf {\bibinfo {volume} {139}},\ \href
  {https://doi.org/10.1063/1.4827025} {10.1063/1.4827025} (\bibinfo {year}
  {2013})\BibitemShut {NoStop}%
\bibitem [{\citenamefont {Yu}\ \emph {et~al.}(2019)\citenamefont {Yu},
  \citenamefont {Hendrickson},\ and\ \citenamefont {Yue}}]{Yu2019}%
  \BibitemOpen
  \bibfield  {author} {\bibinfo {author} {\bibfnamefont {X.}~\bibnamefont
  {Yu}}, \bibinfo {author} {\bibfnamefont {K.}~\bibnamefont {Hendrickson}},\
  and\ \bibinfo {author} {\bibfnamefont {D.~K.}\ \bibnamefont {Yue}},\
  }\bibfield  {title} {\bibinfo {title} {{Scale separation and dependence of
  entrainment bubble-size distribution in free-surface turbulence}},\ }\href
  {https://doi.org/10.1017/jfm.2019.986} {\bibfield  {journal} {\bibinfo
  {journal} {Journal of Fluid Mechanics}\ }\textbf {\bibinfo {volume} {885}},\
  \bibinfo {pages} {R2} (\bibinfo {year} {2019})}\BibitemShut {NoStop}%
\bibitem [{\citenamefont {Mukherjee}\ \emph {et~al.}(2019)\citenamefont
  {Mukherjee}, \citenamefont {Safdari}, \citenamefont {Shardt}, \citenamefont
  {Kenjeres}, \citenamefont {den Akker}, \citenamefont {Kenjere{\v{s}}},\ and\
  \citenamefont {{Van Den Akker}}}]{Mukherjee2019}%
  \BibitemOpen
  \bibfield  {author} {\bibinfo {author} {\bibfnamefont {S.}~\bibnamefont
  {Mukherjee}}, \bibinfo {author} {\bibfnamefont {A.}~\bibnamefont {Safdari}},
  \bibinfo {author} {\bibfnamefont {O.}~\bibnamefont {Shardt}}, \bibinfo
  {author} {\bibfnamefont {S.}~\bibnamefont {Kenjeres}}, \bibinfo {author}
  {\bibfnamefont {H.~E. A.~V.}\ \bibnamefont {den Akker}}, \bibinfo {author}
  {\bibfnamefont {S.}~\bibnamefont {Kenjere{\v{s}}}},\ and\ \bibinfo {author}
  {\bibfnamefont {H.~E.}\ \bibnamefont {{Van Den Akker}}},\ }\bibfield  {title}
  {\bibinfo {title} {{Droplet-Turbulence interactions and quasi-equilibrium
  dynamics in turbulent emulsions}},\ }\href
  {https://doi.org/10.1017/jfm.2019.654} {\bibfield  {journal} {\bibinfo
  {journal} {Journal of Fluid Mechanics}\ }\textbf {\bibinfo {volume} {878}},\
  \bibinfo {pages} {221} (\bibinfo {year} {2019})},\ \Eprint
  {https://arxiv.org/abs/1902.09929} {arXiv:1902.09929} \BibitemShut {NoStop}%
\bibitem [{\citenamefont {Perrard}\ \emph {et~al.}(2021)\citenamefont
  {Perrard}, \citenamefont {Rivi{\`e}re}, \citenamefont {Mostert},\ and\
  \citenamefont {Deike}}]{Perrard2020}%
  \BibitemOpen
  \bibfield  {author} {\bibinfo {author} {\bibfnamefont {S.}~\bibnamefont
  {Perrard}}, \bibinfo {author} {\bibfnamefont {A.}~\bibnamefont
  {Rivi{\`e}re}}, \bibinfo {author} {\bibfnamefont {W.}~\bibnamefont
  {Mostert}},\ and\ \bibinfo {author} {\bibfnamefont {L.}~\bibnamefont
  {Deike}},\ }\bibfield  {title} {\bibinfo {title} {Bubble deformation by a
  turbulent flow},\ }\href@noop {} {\bibfield  {journal} {\bibinfo  {journal}
  {Journal of Fluid Mechanics}\ }\textbf {\bibinfo {volume} {920}} (\bibinfo
  {year} {2021})}\BibitemShut {NoStop}%
\bibitem [{\citenamefont {Rivi{\`{e}}re}\ \emph {et~al.}(2021)\citenamefont
  {Rivi{\`{e}}re}, \citenamefont {Mostert}, \citenamefont {Perrard},\ and\
  \citenamefont {Deike}}]{Riviere2021}%
  \BibitemOpen
  \bibfield  {author} {\bibinfo {author} {\bibfnamefont {A.}~\bibnamefont
  {Rivi{\`{e}}re}}, \bibinfo {author} {\bibfnamefont {W.}~\bibnamefont
  {Mostert}}, \bibinfo {author} {\bibfnamefont {S.}~\bibnamefont {Perrard}},\
  and\ \bibinfo {author} {\bibfnamefont {L.}~\bibnamefont {Deike}},\ }\bibfield
   {title} {\bibinfo {title} {{Sub-Hinze scale bubble production in turbulent
  bubble break-up}},\ }\href {https://doi.org/10.1017/jfm.2021.243} {\bibfield
  {journal} {\bibinfo  {journal} {Journal of Fluid Mechanics}\ }\textbf
  {\bibinfo {volume} {917}},\ \bibinfo {pages} {A40} (\bibinfo {year}
  {2021})}\BibitemShut {NoStop}%
\bibitem [{\citenamefont {Yi}\ \emph {et~al.}(2021)\citenamefont {Yi},
  \citenamefont {Toschi},\ and\ \citenamefont {Sun}}]{Yi2021}%
  \BibitemOpen
  \bibfield  {author} {\bibinfo {author} {\bibfnamefont {L.}~\bibnamefont
  {Yi}}, \bibinfo {author} {\bibfnamefont {F.}~\bibnamefont {Toschi}},\ and\
  \bibinfo {author} {\bibfnamefont {C.}~\bibnamefont {Sun}},\ }\bibfield
  {title} {\bibinfo {title} {{Global and local statistics in turbulent
  emulsions}},\ }\href {https://doi.org/10.1017/jfm.2020.1118} {\bibfield
  {journal} {\bibinfo  {journal} {Journal of Fluid Mechanics}\ }\textbf
  {\bibinfo {volume} {912}},\ \bibinfo {pages} {A13} (\bibinfo {year}
  {2021})},\ \Eprint {https://arxiv.org/abs/2011.00963} {arXiv:2011.00963}
  \BibitemShut {NoStop}%
\bibitem [{\citenamefont {Kolmogorov}(1949)}]{Kolmogorov1949}%
  \BibitemOpen
  \bibfield  {author} {\bibinfo {author} {\bibfnamefont {A.}~\bibnamefont
  {Kolmogorov}},\ }\bibfield  {title} {\bibinfo {title} {{On the breakage of
  drops in a turbulent flow}},\ }\href@noop {} {\bibfield  {journal} {\bibinfo
  {journal} {Dokl. Akad. Navk. SSSR}\ }\textbf {\bibinfo {volume} {66}},\
  \bibinfo {pages} {825} (\bibinfo {year} {1949})}\BibitemShut {NoStop}%
\bibitem [{\citenamefont {Hinze}(1955)}]{Hinze1955}%
  \BibitemOpen
  \bibfield  {author} {\bibinfo {author} {\bibfnamefont {J.~O.}\ \bibnamefont
  {Hinze}},\ }\bibfield  {title} {\bibinfo {title} {{Fundamentals of the
  hydrodynamic mechanism of splitting in dispersion processes}},\ }\href
  {https://doi.org/10.1002/aic.690010303} {\bibfield  {journal} {\bibinfo
  {journal} {AIChE Journal}\ }\textbf {\bibinfo {volume} {1}},\ \bibinfo
  {pages} {289} (\bibinfo {year} {1955})}\BibitemShut {NoStop}%
\bibitem [{\citenamefont {Qi}\ \emph {et~al.}(2022)\citenamefont {Qi},
  \citenamefont {Tan}, \citenamefont {Corbitt}, \citenamefont {Urbanik},
  \citenamefont {Salibindla},\ and\ \citenamefont {Ni}}]{qi2022fragmentation}%
  \BibitemOpen
  \bibfield  {author} {\bibinfo {author} {\bibfnamefont {Y.}~\bibnamefont
  {Qi}}, \bibinfo {author} {\bibfnamefont {S.}~\bibnamefont {Tan}}, \bibinfo
  {author} {\bibfnamefont {N.}~\bibnamefont {Corbitt}}, \bibinfo {author}
  {\bibfnamefont {C.}~\bibnamefont {Urbanik}}, \bibinfo {author} {\bibfnamefont
  {A.~K.}\ \bibnamefont {Salibindla}},\ and\ \bibinfo {author} {\bibfnamefont
  {R.}~\bibnamefont {Ni}},\ }\bibfield  {title} {\bibinfo {title}
  {Fragmentation in turbulence by small eddies},\ }\href@noop {} {\bibfield
  {journal} {\bibinfo  {journal} {Nature Communications}\ }\textbf {\bibinfo
  {volume} {13}},\ \bibinfo {pages} {1} (\bibinfo {year} {2022})}\BibitemShut
  {NoStop}%
\bibitem [{\citenamefont {Fuster}\ and\ \citenamefont
  {Rossi}(2021)}]{fuster2021vortex}%
  \BibitemOpen
  \bibfield  {author} {\bibinfo {author} {\bibfnamefont {D.}~\bibnamefont
  {Fuster}}\ and\ \bibinfo {author} {\bibfnamefont {M.}~\bibnamefont {Rossi}},\
  }\bibfield  {title} {\bibinfo {title} {Vortex-interface interactions in
  two-dimensional flows},\ }\href@noop {} {\bibfield  {journal} {\bibinfo
  {journal} {International Journal of Multiphase Flow}\ }\textbf {\bibinfo
  {volume} {143}},\ \bibinfo {pages} {103757} (\bibinfo {year}
  {2021})}\BibitemShut {NoStop}%
\bibitem [{\citenamefont {Kolmogorov}(1991)}]{Kolmogorov1941}%
  \BibitemOpen
  \bibfield  {author} {\bibinfo {author} {\bibfnamefont {A.~N.}\ \bibnamefont
  {Kolmogorov}},\ }\bibfield  {title} {\bibinfo {title} {{The Local Structure
  of Turbulence in Incompressible Viscous Fluid for Very Large Reynolds
  Numbers}},\ }\href {https://doi.org/10.1098/rspa.1991.0075} {\bibfield
  {journal} {\bibinfo  {journal} {Proceedings of the Royal Society A:
  Mathematical, Physical and Engineering Sciences}\ }\textbf {\bibinfo {volume}
  {434}},\ \bibinfo {pages} {9} (\bibinfo {year} {1991})}\BibitemShut {NoStop}%
\bibitem [{\citenamefont {Deane}\ and\ \citenamefont
  {Stokes}(2002)}]{Deane2002}%
  \BibitemOpen
  \bibfield  {author} {\bibinfo {author} {\bibfnamefont {G.~B.}\ \bibnamefont
  {Deane}}\ and\ \bibinfo {author} {\bibfnamefont {M.~D.}\ \bibnamefont
  {Stokes}},\ }\bibfield  {title} {\bibinfo {title} {{Scale dependence of
  bubble creation mechanisms in breaking waves}},\ }\href
  {https://doi.org/10.1038/nature00967} {\bibfield  {journal} {\bibinfo
  {journal} {Nature}\ }\textbf {\bibinfo {volume} {418}},\ \bibinfo {pages}
  {839} (\bibinfo {year} {2002})}\BibitemShut {NoStop}%
\bibitem [{\citenamefont {Blenkinsopp}\ and\ \citenamefont
  {Chaplin}(2010)}]{blenkinsopp2010bubble}%
  \BibitemOpen
  \bibfield  {author} {\bibinfo {author} {\bibfnamefont {C.~E.}\ \bibnamefont
  {Blenkinsopp}}\ and\ \bibinfo {author} {\bibfnamefont {J.~R.}\ \bibnamefont
  {Chaplin}},\ }\bibfield  {title} {\bibinfo {title} {Bubble size measurements
  in breaking waves using optical fiber phase detection probes},\ }\href@noop
  {} {\bibfield  {journal} {\bibinfo  {journal} {IEEE Journal of Oceanic
  Engineering}\ }\textbf {\bibinfo {volume} {35}},\ \bibinfo {pages} {388}
  (\bibinfo {year} {2010})}\BibitemShut {NoStop}%
\bibitem [{\citenamefont {Wang}\ \emph {et~al.}(2016)\citenamefont {Wang},
  \citenamefont {Yang},\ and\ \citenamefont {Stern}}]{wang2016high}%
  \BibitemOpen
  \bibfield  {author} {\bibinfo {author} {\bibfnamefont {Z.}~\bibnamefont
  {Wang}}, \bibinfo {author} {\bibfnamefont {J.}~\bibnamefont {Yang}},\ and\
  \bibinfo {author} {\bibfnamefont {F.}~\bibnamefont {Stern}},\ }\bibfield
  {title} {\bibinfo {title} {High-fidelity simulations of bubble, droplet and
  spray formation in breaking waves},\ }\href@noop {} {\bibfield  {journal}
  {\bibinfo  {journal} {Journal of Fluid Mechanics}\ }\textbf {\bibinfo
  {volume} {792}},\ \bibinfo {pages} {307} (\bibinfo {year}
  {2016})}\BibitemShut {NoStop}%
\bibitem [{\citenamefont {Deike}\ \emph {et~al.}(2016)\citenamefont {Deike},
  \citenamefont {Melville},\ and\ \citenamefont {Popinet}}]{Deike:2016ir}%
  \BibitemOpen
  \bibfield  {author} {\bibinfo {author} {\bibfnamefont {L.}~\bibnamefont
  {Deike}}, \bibinfo {author} {\bibfnamefont {W.~K.}\ \bibnamefont
  {Melville}},\ and\ \bibinfo {author} {\bibfnamefont {S.}~\bibnamefont
  {Popinet}},\ }\bibfield  {title} {\bibinfo {title} {{Air entrainment and
  bubble statistics in breaking waves}},\ }\href@noop {} {\bibfield  {journal}
  {\bibinfo  {journal} {J. Fluid Mech.}\ }\textbf {\bibinfo {volume} {801}},\
  \bibinfo {pages} {91} (\bibinfo {year} {2016})}\BibitemShut {NoStop}%
\bibitem [{\citenamefont {Chan}\ \emph {et~al.}(2021)\citenamefont {Chan},
  \citenamefont {Johnson}, \citenamefont {Moin},\ and\ \citenamefont
  {Urzay}}]{Chan2020a}%
  \BibitemOpen
  \bibfield  {author} {\bibinfo {author} {\bibfnamefont {W.~H.~R.}\
  \bibnamefont {Chan}}, \bibinfo {author} {\bibfnamefont {P.~L.}\ \bibnamefont
  {Johnson}}, \bibinfo {author} {\bibfnamefont {P.}~\bibnamefont {Moin}},\ and\
  \bibinfo {author} {\bibfnamefont {J.}~\bibnamefont {Urzay}},\ }\bibfield
  {title} {\bibinfo {title} {{The turbulent bubble break-up cascade. Part 2.
  Numerical simulations of breaking waves}},\ }\href
  {https://doi.org/10.1017/jfm.2020.1084} {\bibfield  {journal} {\bibinfo
  {journal} {Journal of Fluid Mechanics}\ }\textbf {\bibinfo {volume} {912}},\
  \bibinfo {pages} {A43} (\bibinfo {year} {2021})},\ \Eprint
  {https://arxiv.org/abs/2009.04804} {arXiv:2009.04804} \BibitemShut {NoStop}%
\bibitem [{\citenamefont {MART{\'I}NEZ-BAZ{\'A}N}\ \emph
  {et~al.}(1999)\citenamefont {MART{\'I}NEZ-BAZ{\'A}N}, \citenamefont
  {Montanes},\ and\ \citenamefont {Lasheras}}]{martinez1999breakup}%
  \BibitemOpen
  \bibfield  {author} {\bibinfo {author} {\bibfnamefont {C.}~\bibnamefont
  {MART{\'I}NEZ-BAZ{\'A}N}}, \bibinfo {author} {\bibfnamefont {J.}~\bibnamefont
  {Montanes}},\ and\ \bibinfo {author} {\bibfnamefont {J.~C.}\ \bibnamefont
  {Lasheras}},\ }\bibfield  {title} {\bibinfo {title} {On the breakup of an air
  bubble injected into a fully developed turbulent flow. part 1. breakup
  frequency},\ }\href@noop {} {\bibfield  {journal} {\bibinfo  {journal}
  {Journal of Fluid Mechanics}\ }\textbf {\bibinfo {volume} {401}},\ \bibinfo
  {pages} {157} (\bibinfo {year} {1999})}\BibitemShut {NoStop}%
\bibitem [{\citenamefont {Benzi}\ \emph {et~al.}(1984)\citenamefont {Benzi},
  \citenamefont {Paladin}, \citenamefont {Parisi},\ and\ \citenamefont
  {Vulpiani}}]{benzi1984multifractal}%
  \BibitemOpen
  \bibfield  {author} {\bibinfo {author} {\bibfnamefont {R.}~\bibnamefont
  {Benzi}}, \bibinfo {author} {\bibfnamefont {G.}~\bibnamefont {Paladin}},
  \bibinfo {author} {\bibfnamefont {G.}~\bibnamefont {Parisi}},\ and\ \bibinfo
  {author} {\bibfnamefont {A.}~\bibnamefont {Vulpiani}},\ }\bibfield  {title}
  {\bibinfo {title} {On the multifractal nature of fully developed turbulence
  and chaotic systems},\ }\href@noop {} {\bibfield  {journal} {\bibinfo
  {journal} {Journal of Physics A: Mathematical and General}\ }\textbf
  {\bibinfo {volume} {17}},\ \bibinfo {pages} {3521} (\bibinfo {year}
  {1984})}\BibitemShut {NoStop}%
\bibitem [{\citenamefont {Meneveau}\ and\ \citenamefont
  {Sreenivasan}(1987)}]{meneveau1987simple}%
  \BibitemOpen
  \bibfield  {author} {\bibinfo {author} {\bibfnamefont {C.}~\bibnamefont
  {Meneveau}}\ and\ \bibinfo {author} {\bibfnamefont {K.}~\bibnamefont
  {Sreenivasan}},\ }\bibfield  {title} {\bibinfo {title} {Simple multifractal
  cascade model for fully developed turbulence},\ }\href@noop {} {\bibfield
  {journal} {\bibinfo  {journal} {Physical review letters}\ }\textbf {\bibinfo
  {volume} {59}},\ \bibinfo {pages} {1424} (\bibinfo {year}
  {1987})}\BibitemShut {NoStop}%
\bibitem [{\citenamefont {Boffetta}\ \emph {et~al.}(2008)\citenamefont
  {Boffetta}, \citenamefont {Mazzino},\ and\ \citenamefont
  {Vulpiani}}]{Boffetta2008}%
  \BibitemOpen
  \bibfield  {author} {\bibinfo {author} {\bibfnamefont {G.}~\bibnamefont
  {Boffetta}}, \bibinfo {author} {\bibfnamefont {A.}~\bibnamefont {Mazzino}},\
  and\ \bibinfo {author} {\bibfnamefont {A.}~\bibnamefont {Vulpiani}},\
  }\bibfield  {title} {\bibinfo {title} {{Twenty-five years of multifractals in
  fully developed turbulence: A tribute to Giovanni Paladin}},\ }\bibfield
  {journal} {\bibinfo  {journal} {Journal of Physics A: Mathematical and
  Theoretical}\ }\textbf {\bibinfo {volume} {41}},\ \href
  {https://doi.org/10.1088/1751-8113/41/36/363001}
  {10.1088/1751-8113/41/36/363001} (\bibinfo {year} {2008}),\ \Eprint
  {https://arxiv.org/abs/0809.0196} {arXiv:0809.0196} \BibitemShut {NoStop}%
\bibitem [{\citenamefont {Frisch}(1995)}]{Frisch1995a}%
  \BibitemOpen
  \bibfield  {author} {\bibinfo {author} {\bibfnamefont {U.}~\bibnamefont
  {Frisch}},\ }\href {https://doi.org/10.1017/CBO9781139170666} {\emph
  {\bibinfo {title} {{Turbulence}}}}\ (\bibinfo  {publisher} {Cambridge
  University Press},\ \bibinfo {year} {1995})\BibitemShut {NoStop}%
\bibitem [{\citenamefont {Pope}(2009)}]{Pope2009}%
  \BibitemOpen
  \bibfield  {author} {\bibinfo {author} {\bibfnamefont {S.}~\bibnamefont
  {Pope}},\ }\href@noop {} {\emph {\bibinfo {title} {{Turbulent Flows}}}},\
  \bibinfo {edition} {sixth}\ ed.\ (\bibinfo  {publisher} {Cambridge University
  Press},\ \bibinfo {year} {2009})\BibitemShut {NoStop}%
\bibitem [{\citenamefont {Alexakis}\ and\ \citenamefont
  {Biferale}(2018{\natexlab{a}})}]{alexakis2018cascades}%
  \BibitemOpen
  \bibfield  {author} {\bibinfo {author} {\bibfnamefont {A.}~\bibnamefont
  {Alexakis}}\ and\ \bibinfo {author} {\bibfnamefont {L.}~\bibnamefont
  {Biferale}},\ }\bibfield  {title} {\bibinfo {title} {Cascades and transitions
  in turbulent flows},\ }\href@noop {} {\bibfield  {journal} {\bibinfo
  {journal} {Physics Reports}\ }\textbf {\bibinfo {volume} {767}},\ \bibinfo
  {pages} {1} (\bibinfo {year} {2018}{\natexlab{a}})}\BibitemShut {NoStop}%
\bibitem [{\citenamefont {Riviere}\ \emph {et~al.}(2021)\citenamefont
  {Riviere}, \citenamefont {Ruth}, \citenamefont {Mostert}, \citenamefont
  {Deike},\ and\ \citenamefont {Perrard}}]{Riviere2021a}%
  \BibitemOpen
  \bibfield  {author} {\bibinfo {author} {\bibfnamefont {A.}~\bibnamefont
  {Riviere}}, \bibinfo {author} {\bibfnamefont {D.}~\bibnamefont {Ruth}},
  \bibinfo {author} {\bibfnamefont {W.}~\bibnamefont {Mostert}}, \bibinfo
  {author} {\bibfnamefont {L.}~\bibnamefont {Deike}},\ and\ \bibinfo {author}
  {\bibfnamefont {S.}~\bibnamefont {Perrard}},\ }\bibfield  {title} {\bibinfo
  {title} {Capillary driven fragmentation of large gas bubbles in turbulence},\
  }\href@noop {} {\bibfield  {journal} {\bibinfo  {journal} {arXiv preprint
  arXiv:2112.06480}\ } (\bibinfo {year} {2021})}\BibitemShut {NoStop}%
\bibitem [{\citenamefont {Buaria}\ and\ \citenamefont
  {Pumir}(2022)}]{buaria2022vorticity}%
  \BibitemOpen
  \bibfield  {author} {\bibinfo {author} {\bibfnamefont {D.}~\bibnamefont
  {Buaria}}\ and\ \bibinfo {author} {\bibfnamefont {A.}~\bibnamefont {Pumir}},\
  }\bibfield  {title} {\bibinfo {title} {Vorticity-strain rate dynamics and the
  smallest scales of turbulence},\ }\href@noop {} {\bibfield  {journal}
  {\bibinfo  {journal} {Physical Review Letters}\ }\textbf {\bibinfo {volume}
  {128}},\ \bibinfo {pages} {094501} (\bibinfo {year} {2022})}\BibitemShut
  {NoStop}%
\bibitem [{\citenamefont {Kolmogorov}(1962)}]{Kolmogorov1962}%
  \BibitemOpen
  \bibfield  {author} {\bibinfo {author} {\bibfnamefont {A.~N.}\ \bibnamefont
  {Kolmogorov}},\ }\bibfield  {title} {\bibinfo {title} {{A refinement of
  previous hypotheses concerning the local structure of turbulence in a viscous
  incompressible fluid at high Reynolds number}},\ }\href
  {https://doi.org/10.1017/S0022112062000518} {\bibfield  {journal} {\bibinfo
  {journal} {Journal of Fluid Mechanics}\ }\textbf {\bibinfo {volume} {13}},\
  \bibinfo {pages} {82} (\bibinfo {year} {1962})}\BibitemShut {NoStop}%
\bibitem [{\citenamefont {Dubrulle}(2019)}]{Dubrulle2019}%
  \BibitemOpen
  \bibfield  {author} {\bibinfo {author} {\bibfnamefont {B.}~\bibnamefont
  {Dubrulle}},\ }\bibfield  {title} {\bibinfo {title} {{Beyond Kolmogorov
  cascades}},\ }\href {https://doi.org/10.1017/jfm.2019.98} {\bibfield
  {journal} {\bibinfo  {journal} {Journal of Fluid Mechanics}\ }\textbf
  {\bibinfo {volume} {867}},\ \bibinfo {pages} {P1} (\bibinfo {year}
  {2019})}\BibitemShut {NoStop}%
\bibitem [{\citenamefont {Buzzicotti}\ \emph {et~al.}(2020)\citenamefont
  {Buzzicotti}, \citenamefont {Biferale},\ and\ \citenamefont
  {Toschi}}]{Buzzicotti2020}%
  \BibitemOpen
  \bibfield  {author} {\bibinfo {author} {\bibfnamefont {M.}~\bibnamefont
  {Buzzicotti}}, \bibinfo {author} {\bibfnamefont {L.}~\bibnamefont
  {Biferale}},\ and\ \bibinfo {author} {\bibfnamefont {F.}~\bibnamefont
  {Toschi}},\ }\bibfield  {title} {\bibinfo {title} {Statistical properties of
  turbulence in the presence of a smart small-scale control},\ }\href@noop {}
  {\bibfield  {journal} {\bibinfo  {journal} {Physical Review Letters}\
  }\textbf {\bibinfo {volume} {124}},\ \bibinfo {pages} {084504} (\bibinfo
  {year} {2020})}\BibitemShut {NoStop}%
\bibitem [{\citenamefont {Crialesi-Esposito}\ \emph {et~al.}(2022)\citenamefont
  {Crialesi-Esposito}, \citenamefont {Rosti}, \citenamefont {Chibbaro},\ and\
  \citenamefont {Brandt}}]{crialesi2022}%
  \BibitemOpen
  \bibfield  {author} {\bibinfo {author} {\bibfnamefont {M.}~\bibnamefont
  {Crialesi-Esposito}}, \bibinfo {author} {\bibfnamefont {M.~E.}\ \bibnamefont
  {Rosti}}, \bibinfo {author} {\bibfnamefont {S.}~\bibnamefont {Chibbaro}},\
  and\ \bibinfo {author} {\bibfnamefont {L.}~\bibnamefont {Brandt}},\
  }\bibfield  {title} {\bibinfo {title} {Modulation of homogeneous and
  isotropic turbulence in emulsions},\ }\href
  {https://doi.org/10.1017/jfm.2022.179} {\bibfield  {journal} {\bibinfo
  {journal} {Journal of Fluid Mechanics}\ }\textbf {\bibinfo {volume} {940}},\
  \bibinfo {pages} {A19} (\bibinfo {year} {2022})}\BibitemShut {NoStop}%
\bibitem [{\citenamefont {Podvigina}\ and\ \citenamefont
  {Pouquet}(1994)}]{Podvigina1994}%
  \BibitemOpen
  \bibfield  {author} {\bibinfo {author} {\bibfnamefont {O.}~\bibnamefont
  {Podvigina}}\ and\ \bibinfo {author} {\bibfnamefont {A.}~\bibnamefont
  {Pouquet}},\ }\bibfield  {title} {\bibinfo {title} {{On the non-linear
  stability of the 1:1:1 ABC flow}},\ }\href
  {https://doi.org/10.1016/0167-2789(94)00031-X} {\bibfield  {journal}
  {\bibinfo  {journal} {Physica D: Nonlinear Phenomena}\ }\textbf {\bibinfo
  {volume} {75}},\ \bibinfo {pages} {471} (\bibinfo {year} {1994})}\BibitemShut
  {NoStop}%
\bibitem [{\citenamefont {Mininni}\ \emph {et~al.}(2006)\citenamefont
  {Mininni}, \citenamefont {Alexakis},\ and\ \citenamefont
  {Pouquet}}]{Mininni2006}%
  \BibitemOpen
  \bibfield  {author} {\bibinfo {author} {\bibfnamefont {P.~D.}\ \bibnamefont
  {Mininni}}, \bibinfo {author} {\bibfnamefont {A.}~\bibnamefont {Alexakis}},\
  and\ \bibinfo {author} {\bibfnamefont {A.}~\bibnamefont {Pouquet}},\
  }\bibfield  {title} {\bibinfo {title} {{Large-scale flow effects, energy
  transfer, and self-similarity on turbulence}},\ }\href
  {https://doi.org/10.1103/PhysRevE.74.016303} {\bibfield  {journal} {\bibinfo
  {journal} {Physical Review E - Statistical, Nonlinear, and Soft Matter
  Physics}\ }\textbf {\bibinfo {volume} {74}},\ \bibinfo {pages} {1} (\bibinfo
  {year} {2006})}\BibitemShut {NoStop}%
\bibitem [{\citenamefont {Rosti}\ \emph {et~al.}(2020)\citenamefont {Rosti},
  \citenamefont {Ge}, \citenamefont {Jain}, \citenamefont {Dodd},\ and\
  \citenamefont {Brandt}}]{Rosti2020}%
  \BibitemOpen
  \bibfield  {author} {\bibinfo {author} {\bibfnamefont {M.~E.}\ \bibnamefont
  {Rosti}}, \bibinfo {author} {\bibfnamefont {Z.}~\bibnamefont {Ge}}, \bibinfo
  {author} {\bibfnamefont {S.~S.}\ \bibnamefont {Jain}}, \bibinfo {author}
  {\bibfnamefont {M.~S.}\ \bibnamefont {Dodd}},\ and\ \bibinfo {author}
  {\bibfnamefont {L.}~\bibnamefont {Brandt}},\ }\bibfield  {title} {\bibinfo
  {title} {{Droplets in homogeneous shear turbulence}},\ }\href
  {https://doi.org/10.1017/jfm.2019.581} {\bibfield  {journal} {\bibinfo
  {journal} {J. Fluid Mech}\ }\textbf {\bibinfo {volume} {876}},\ \bibinfo
  {pages} {962} (\bibinfo {year} {2020})}\BibitemShut {NoStop}%
\bibitem [{\citenamefont {Costa}(2018)}]{Costa2018}%
  \BibitemOpen
  \bibfield  {author} {\bibinfo {author} {\bibfnamefont {P.}~\bibnamefont
  {Costa}},\ }\bibfield  {title} {\bibinfo {title} {{A FFT-based
  finite-difference solver for massively-parallel direct numerical simulations
  of turbulent flows}},\ }\href {https://doi.org/10.1016/j.camwa.2018.07.034}
  {\bibfield  {journal} {\bibinfo  {journal} {Computers and Mathematics with
  Applications}\ }\textbf {\bibinfo {volume} {76}},\ \bibinfo {pages} {1853}
  (\bibinfo {year} {2018})},\ \Eprint
  {https://arxiv.org/abs/arXiv:1802.10323v3} {arXiv:arXiv:1802.10323v3}
  \BibitemShut {NoStop}%
\bibitem [{\citenamefont {Ii}\ \emph {et~al.}(2012)\citenamefont {Ii},
  \citenamefont {Sugiyama}, \citenamefont {Takeuchi}, \citenamefont {Takagi},
  \citenamefont {Matsumoto},\ and\ \citenamefont {Xiao}}]{Ii2012}%
  \BibitemOpen
  \bibfield  {author} {\bibinfo {author} {\bibfnamefont {S.}~\bibnamefont
  {Ii}}, \bibinfo {author} {\bibfnamefont {K.}~\bibnamefont {Sugiyama}},
  \bibinfo {author} {\bibfnamefont {S.}~\bibnamefont {Takeuchi}}, \bibinfo
  {author} {\bibfnamefont {S.}~\bibnamefont {Takagi}}, \bibinfo {author}
  {\bibfnamefont {Y.}~\bibnamefont {Matsumoto}},\ and\ \bibinfo {author}
  {\bibfnamefont {F.}~\bibnamefont {Xiao}},\ }\bibfield  {title} {\bibinfo
  {title} {{An interface capturing method with a continuous function: The THINC
  method with multi-dimensional reconstruction}},\ }\href
  {https://doi.org/10.1016/j.jcp.2011.11.038} {\bibfield  {journal} {\bibinfo
  {journal} {Journal of Computational Physics}\ }\textbf {\bibinfo {volume}
  {231}},\ \bibinfo {pages} {2328} (\bibinfo {year} {2012})}\BibitemShut
  {NoStop}%
\bibitem [{\citenamefont {Komrakova}\ \emph {et~al.}(2015)\citenamefont
  {Komrakova}, \citenamefont {Eskin},\ and\ \citenamefont
  {Derksen}}]{Komrakova2015}%
  \BibitemOpen
  \bibfield  {author} {\bibinfo {author} {\bibfnamefont {A.~E.}\ \bibnamefont
  {Komrakova}}, \bibinfo {author} {\bibfnamefont {D.}~\bibnamefont {Eskin}},\
  and\ \bibinfo {author} {\bibfnamefont {J.~J.}\ \bibnamefont {Derksen}},\
  }\bibfield  {title} {\bibinfo {title} {{Numerical study of turbulent
  liquid-liquid dispersions}},\ }\href {https://doi.org/10.1002/aic.14821}
  {\bibfield  {journal} {\bibinfo  {journal} {AIChE Journal}\ }\textbf
  {\bibinfo {volume} {61}},\ \bibinfo {pages} {2618} (\bibinfo {year}
  {2015})},\ \Eprint {https://arxiv.org/abs/0201037v1} {arXiv:0201037v1
  [arXiv:physics]} \BibitemShut {NoStop}%
\bibitem [{\citenamefont {Alexakis}\ and\ \citenamefont
  {Biferale}(2018{\natexlab{b}})}]{Alexakis2018}%
  \BibitemOpen
  \bibfield  {author} {\bibinfo {author} {\bibfnamefont {A.}~\bibnamefont
  {Alexakis}}\ and\ \bibinfo {author} {\bibfnamefont {L.}~\bibnamefont
  {Biferale}},\ }\bibfield  {title} {\bibinfo {title} {{Cascades and
  transitions in turbulent flows}},\ }\href
  {https://doi.org/10.1016/j.physrep.2018.08.001} {\bibfield  {journal}
  {\bibinfo  {journal} {Physics Reports}\ }\textbf {\bibinfo {volume}
  {767-769}},\ \bibinfo {pages} {1} (\bibinfo {year}
  {2018}{\natexlab{b}})}\BibitemShut {NoStop}%
\end{thebibliography}%

\section*{Methods}
\subsection*{Numerical simulation}
We study emulsions in homogeneous and isotropic turbulence by means of direct numerical simulations.
The problem is described by the one-fluid formulation of the multiphase Navier-Stokes equation:
\begin{equation}
\rho(\partial_t u_i+u_j\partial_j u_i) = -\partial_i p + \partial_i \left[ \mu (\partial_i u_j + \partial_j u_i) \right] +f^\sigma_i +f^T_i~,
\label{eq:ns}
\end{equation}
where $u_i$ is the velocity field, $p$ is the pressure, $\mu$ is the flow viscosity, and $\rho$ the fluid density. The  term $f^\sigma_i=\sigma \xi \delta_s n_i$ represents the surface tension forces, where $\delta_s$ is a Dirac delta function that concentrate the term action at the surface, with $\xi$ and $n_i$  the interface curvature  and normal vector.
The term $f^T_i$ is the large scale forcing, used to sustain turbulence throughout the simulation box of size $L=2\pi$. The forcing is the Arnold-Beltrami-Childress (ABC) \cite{Podvigina1994,Mininni2006}, implemented as: 

\begin{align}
f^T_x &= A\: sin \:\kappa_0 z + C\: cos \:\kappa_0 y \\
f^T_y &= B\: sin \:\kappa_0 x + A\: cos \:\kappa_0 z \\
f^T_z &= C\: sin \:\kappa_0 y + B\: cos \:\kappa_0 x.
\end{align}

In order to avoid large-scale coalescence effects, turbulence is forced at $k_0=2\pi/\mathcal{L}$, where $\mathcal{L}$ is the injection scale. For all cases presented, $A=B=C=1$. 

The algorithm used to solve \Cref{eq:ns} is described in \cite{Rosti2020}, while further details on the direct FFT solver used to solve the pressure Poisson equation are provided in \cite{Costa2018}. The interface is captured with the algebraic Volume of Fluid (VoF) method MTHINC from \cite{Ii2012}.

\subsection*{Case setup}
The cases discussed in this work are presented in \Cref{tab:testMat}.
All simulations are performed on a box domain of size $2\pi$, with turbulence forced at $k_0=2$ in order to avoid coalescence  induced by the large scale dynamics \cite{Komrakova2015}. The simulation box is discretized with $512^3$ grid points. The total simulation time reported in the table is quantified in terms of large-eddy turnover times, $N_\mathcal{T}=t_{tot}/\mathcal{T}_\mathcal{L}$, with $\mathcal{T}_\mathcal{L}=\mathcal{L} u_{rms}$ (see \cite{Mininni2006}), where $u_{rms}$ is the velocity fluctuation root-mean-square value.  
The dispersed phase is initialised on a fully-developed single-phase turbulent field at Taylor-scale Reynolds number $Re_\lambda=137$. We explore different
large-scale Weber number, the ratio between inertial and surface tension forces (i.e. the disperse phase deformability),  defined as  $We_\mathcal{L}=\rho_c\mathcal{L} u_{rms} ^2/\sigma$. Furthermore, we vary the viscosity ratio $\mu_d=\mu_c$ (with $d$ and $c$ being the dispersed and carrier phase) and the dispersed phase volume fraction $\alpha$. For all simulations, $\rho_d=\rho_c=1$.

\begin{table}
	\centering
	\begin{tabular}{c c c c c c}
		&  $\mu_d/\mu_c$ &$We_\mathcal{L}$ &$\sigma$  & $\alpha$ & $N_\mathcal{T}$ \\ 
		\rule{0pt}{4ex}
		SP2 &  - & - & - & - & 136 \\
		\rule{0pt}{4ex}%
		BE1 &  1    & 42.6 & 0.46 & 0.03 & 115\\
		BE2 &  1    & 42.6 & 0.46 & 0.1  & 100\\
		\rule{0pt}{4ex}%
		V11 &  0.01 & 42.6 & 0.46 & 0.03 & 115\\
		V12 &  0.1  & 42.6 & 0.46 & 0.03 & 100\\
		V13 &  10   & 42.6 & 0.46 & 0.03 & 64\\
		V14 &  100  & 42.6 & 0.46 & 0.03 & 60\\
		\rule{0pt}{4ex}%
		V21 &  0.01 & 42.6 & 0.46 & 0.1 & 115\\
		V22 &  0.1  & 42.6 & 0.46 & 0.1 & 100\\
		V23 &  10   & 42.6 & 0.46 & 0.1 & 64\\
		V24 &  100  & 42.6 & 0.46 & 0.1 & 60\\
		\rule{0pt}{4ex}
		C12 &  1    & 42.6 & 0.46 & 0.06 & 100\\
		C13 &  1    & 42.6 & 0.46 & 0.0775 & 100\\
		C14 &  1    & 42.6 & 0.46 & 0.5  & 100\\
		\rule{0pt}{4ex}%
		W11 &  1 & 10.6 & 1.84 &0.03 & 160\\
		W12 & 1 & 21.2 & 0.92 &0.03 & 160\\
		W13 & 1 & 106.5 & 0.184 &0.03 & 100\\

	\end{tabular}
	\caption{Parameter settings for the simulations considered in this study: number of grid points in each direction $N$, viscosity ratio $\mu_d/\mu_c$, Weber number  $We_\mathcal{L}$ with surface tension $\sigma$, volume fraction $\alpha$ and integration time to reach statistical convergence $N_\mathcal{T}$. All simulations are performed with $\mu_c=0.006$ and same ABC forcing. 	Each case is denoted by a letter indicating the parameter which is varied: V for viscosity ratio, C volume fraction and W Weber number. SP are the single-phase flows and BE are configurations which recur in different parameterizations (base emulsions). }
	\label{tab:testMat}
\end{table}

\subsection*{Shell-by-shell energy balance}
Significant insight on the flow dynamics is given by the shell-by-shell energy balance. 
This enables us to quantify the contribution of each term of \Cref{eq:ns} to the energy at each scale $\ell$, or wavenumber $\kappa$ in Fourier space. To derive \Cref{eq:sbs} in the main text, we first perform the Fourier transform (indicated through the symbol ~$\widetilde{\cdot}$~) of  \Cref{eq:ns}, yielding
\begin{equation}
\partial_t \widetilde{u_i} + \widetilde{G_i} = -\mathrm{i}\kappa\widetilde{p/\rho} - \widetilde{V_i} + \widetilde{f^\sigma_i} +\widetilde{f_i^T},
\label{eq:NSspec}
\end{equation}
where $\widetilde{G_i}$ and $\widetilde{V_i}$ are the Fourier transforms of the non-linear and viscous terms. To obtain the energy equation, we multiply \Cref{eq:ns} by $\widetilde{u_i}$ and repeat the same operations for the complex conjugate velocity and sum the two equations. The resulting terms are
$E = \widetilde{u_i}\widetilde{u_i}^*$ (the  kinetic energy in the Fourier space), 
$T = -(\widetilde{G_i}\widetilde{u_i}^* + \widetilde{G_i}^*\widetilde{u_i})$  (the energy transfer due to the non-linear term),
$\mathcal{D} = -(\widetilde{V_i}\widetilde{u_i}^* + \widetilde{V_i}^*\widetilde{u_i})$ (the viscous dissipation),
$\mathcal{S}_{\sigma} = (\widetilde{f_i^\sigma}\widetilde{u_i}^* + \widetilde{f_i^\sigma}^*\widetilde{u_i})$( the work of the surface tension force ) and 
$\mathcal{F} = (\widetilde{f^T_i}\widetilde{u_i}^* + \widetilde{f^T_i}^*\widetilde{u_i})$ (the energy input due to the large-scale forcing).
We finally obtain \Cref{eq:sbs} by performing a shell-integral, \emph{e. g.,} for the surface tension term $\mathcal{S}_{\sigma}(\kappa) = \sum_{\kappa<|\kappa_i|<\kappa+1} \mathcal{S}_{\sigma}(\kappa_i)$.
Further details on the derivation and properties of this equation can be found in \cite{Frisch1995a,Alexakis2018,crialesi2022} 

\end{document}


\newcommand{\lb}[1]{[{\scriptsize\color{red}#1}]}
\newcommand{\LB}[1]{[{\color{magenta}#1}]}
\newcommand{\SC}[1]{[{\color{green}\textbf{#1}}]}
\newcommand{\MC}[1]{[{\color{blue}#1}]}
\newcommand{\wel}{$We_{\mathcal{L}}$}
\newcommand{\mur}{$\mu_d/\mu_c$}
\newcommand{\lra}[1]{\langle #1 \rangle }

\title{Supplementary information.
}%

%
%
%

\maketitle
\definecolor{green}{rgb}{0,0.5,0}
\definecolor{black}{rgb}{0,0,0}
\definecolor{a003}{rgb} {0.0, 0.5, 0.0}
\definecolor{a006}{rgb} {1.0, 0.0, 0.0}
\definecolor{a01}{rgb}  {0.0, 0.0, 1.0}
\definecolor{a02}{rgb}  {0.75, 0.0, 0.75}
\definecolor{a05}{rgb}  {1.0, 0.4980392156862745, 0.054901960784313725}

\definecolor{m001}{rgb} {0.5019607843137255, 0.5019607843137255, 0.0}
\definecolor{m01}{rgb}  {1.0, 0.0784313725490196, 0.5764705882352941}
\definecolor{m1}{rgb}   {0.0, 0.75, 0.75}
\definecolor{m10}{rgb}  {1.0, 0.8431372549019608, 0.0}
\definecolor{m100}{rgb} {0.0, 0.39215686274509803, 0.0}

\definecolor{fam_a}{rgb}     {1.0, 0.0, 0.0}
\definecolor{fam_we}{rgb}    {0.0, 0.5, 0.0}
\definecolor{fam_mu10}{rgb}  {0.75, 0.0, 0.75}
\definecolor{fam_mu3}{rgb}   {1.0, 0.4980392156862745, 0.054901960784313725}

\scalebox{0}{%
\begin{tikzpicture}
    \begin{axis}[hide axis]
     	\addplot [
    	color=black,
    	dashed,
    	line width=0.9pt,
    	forget plot
    	]
    	(0,0);\label{dashed}
    	
    	\addplot [
    	color=black,
    	line width=0.9pt,
    	forget plot
    	]
    	(0,0);\label{conti}
    	
        \addplot [
        color=a003,
        dashed,
        line width=0.9pt,
        forget plot
        ]
        (0,0);\label{alpha003}
        
        \addplot [
        color=a006,
        dashed,
        line width=0.9pt,
        forget plot
        ]
        (0,0);\label{alpha006}
        
        \addplot [
        color=a01,
        dashed,
        line width=0.9pt,
        forget plot
        ]
        (0,0);\label{alpha01}
        
        \addplot [
        color=a02,
        dashed,
        line width=0.9pt,
        forget plot
        ]
        (0,0);\label{alpha02}
        
        \addplot [
        color=a05,
        dashed,
        line width=0.9pt,
        forget plot
        ]
        (0,0);\label{alpha05}
        
        \addplot [
        color=m001,
        dashdotted,
        line width=0.9pt,
        forget plot
        ]
        (0,0);\label{m001}
        
        \addplot [
        color=m01,
        dashdotted,
        line width=0.9pt,
        forget plot
        ]
        (0,0);\label{m01}
        
        \addplot [
        color=m1,
        dashdotted,
        line width=0.9pt,
        forget plot
        ]
        (0,0);\label{m1}
        
        \addplot [
        color=m10,
        dashdotted,
        line width=0.9pt,
        forget plot
        ]
        (0,0);\label{m10}
        
        \addplot [
        color=m100,
        dashdotted,
        line width=0.9pt,
        forget plot
        ]
        (0,0);\label{m100}
        
        \addplot [
        only marks,
        color=fam_a,
        mark=*,
        opacity=0.5,
        forget plot
        ]
        (0,0);\label{fam_a}
        
        \addplot [
        only marks,
        color=fam_we,
        mark=*,
        opacity=0.5,
        forget plot
        ]
        (0,0);\label{fam_we}

        \addplot [
        only marks,
        color=fam_mu3,
        mark=*,
        opacity=0.5,
        forget plot
        ]
        (0,0);\label{fam_mu3}
        
        \addplot [
        only marks,
        color=fam_mu10,
        mark=*,
        opacity=0.5,
        forget plot
        ]
        (0,0);\label{fam_mu10}
        
    \end{axis}
\end{tikzpicture}%
}%
\begin{figure}
	\centering
	\includegraphics[width=0.6\textwidth]{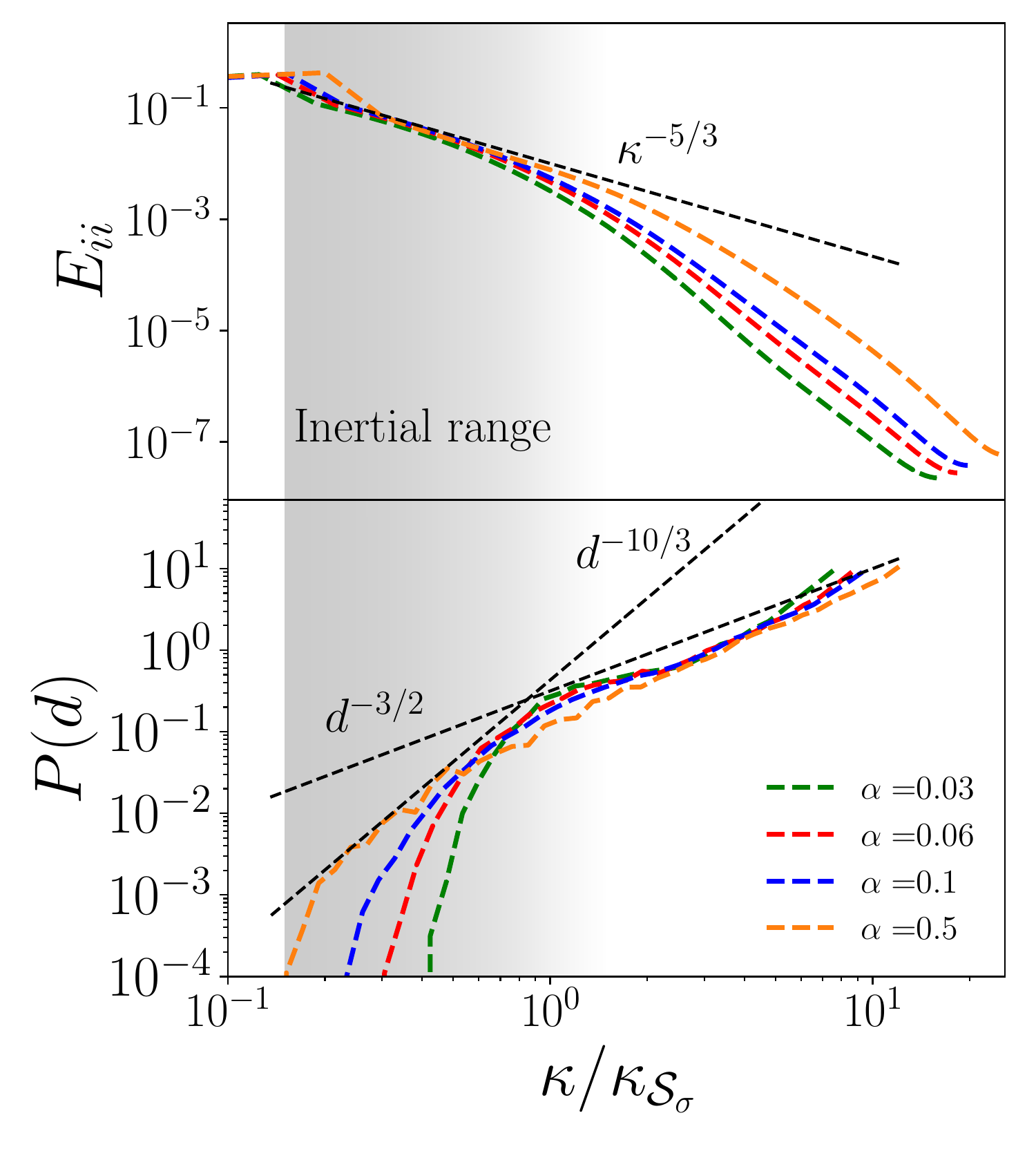}
	\caption{One dimensional energy spectra(top panel) and droplet-size-distribution (bottom panel) for simulations at different volume fractions . Wavenumbers are normalized using $\kappa_{\mathcal{S}_\sigma}$, so that the transition between $-3/2$ and $-10/3$ is clearly visible. In the inertial range a \textit{pseudo-inertial} regime is indicated using a black dashed line for the $-5/3$ scaling (top panel). Also in black dashed lines, droplet-size-distributions power-laws are shown with respective labels (bottom panel).}
\end{figure}

\textbf{Video1:}{Time-sequence of a droplet breakup in turbulence. The simulation is performed at $\mu_d/\mu_c=1$ and $We_\mathcal{L}=42.6$. The volume fraction is set to $\alpha=0.0775$, so that $\kappa_d=2\pi/d\sim 3$. The render shows the droplet interface (white contour) with vorticity magnitude projected onto the simulation box. The panel on the top-left corner shows the surface tension energy transfer function $\mathcal{S}_\sigma$ (see Material and Methods) normalized by its maximum values to improve readability. The dashed-blue line shows the instantaneous value of $\mathcal{S}_\sigma$ (\emph{i.e.} corresponding to the instant rendered), while the time-averaged value in statistical-stationary condition is reported using a black line.
	A red star marker is added as reference for $\kappa_d$. The whole time sequence evolves during approximately one large-eddies turnover time $\mathcal{T}_\mathcal{L}$}